\documentclass[nohyper,notoc]{article} 
\usepackage{epsfig}
\input epsf
\def\beq{\begin{equation}}
\def\eeq{\end{equation}}
\def\be{\begin{equation}}
\def\ee{\end{equation}}
\def\bea{\begin{eqnarray}}
\def\eea{\end{eqnarray}}
\def \lsim{\mathrel{\vcenter
     {\hbox{$<$}\nointerlineskip\hbox{$\sim$}}}}
\def \gsim{\mathrel{\vcenter
     {\hbox{$>$}\nointerlineskip\hbox{$\sim$}}}}
\def\gappeq{\mathrel{\rlap {\raise.5ex\hbox{$>$}}
{\lower.5ex\hbox{$\sim$}}}}
\def\lappeq{\mathrel{\rlap{\raise.5ex\hbox{$<$}}
{\lower.5ex\hbox{$\sim$}}}}

\def\snu{\tilde{\nu}}

\newcommand{\sun}{\Delta m^2_{\rm solar}}
\newcommand{\atm}{\Delta m^2_{\rm atm}}

\begin{document}
\vspace*{-1in}
\renewcommand{\thefootnote}{\fnsymbol{footnote}}
\begin{flushright}
IPPP/02/83\\ DCPT/02/166 
\end{flushright}
\vskip 5pt
\begin{center}
{\Large {\bf From weak-scale
observables to leptogenesis}}
\vskip 25pt
{\bf Sacha Davidson } 
 
\vskip 10pt 
 {\it  IPPP, University of Durham, Durham, DH1 3LE, UK}
\vskip 20pt
{\bf Abstract}
\end{center}
\begin{quotation}
Thermal leptogenesis is an attractive mechanism for
generating the baryon asymmetry of the Universe. However, 
in supersymmetric models, the parameter space is severely restricted
by the gravitino bound on the reheat temperature $T_{RH}$.
For hierarchical
light neutrino masses,  it is shown that  thermal leptogenesis
{\it can} work when $T_{RH} \sim 10^{9} $ GeV.  
The low-energy observable consequences of this
scenario are  $ BR( \tau \rightarrow \ell \gamma)
\sim 10^{-8} - 10^{-9}  $. For higher $T_{RH}$, thermal
leptogenesis works in a larger area of parameter space,
whose observable consequences are more ambiguous.
A parametrisation of the seesaw in terms of
weak-scale inputs is used, so the results are independent of
the texture chosen for the GUT-scale Yukawa matrices.
{\noindent\small 
\vskip 10pt
\noindent
}

\end{quotation}

\vskip 20pt  

\setcounter{footnote}{0}
\renewcommand{\thefootnote}{\arabic{footnote}}


\section{Introduction}

Leptogenesis \cite{Fukugita:1986hr} is an appealing mechanism for producing
the baryon asymmetry of the Universe\cite{Buchmuller:2000wq}. In the 
seesaw model\cite{seesaw},
heavy singlet (``right-handed'') neutrinos $\nu_R$ 
decay out-of-equilibrium,
producing a net lepton asymmetry, which is
reprocessed by Standard Model (SM) $B+L$ violating
processes \cite{Kuzmin:1985mm} into a baryon asymmetry. A natural
and cosmology-independent way to
produce the $\nu_R$ is by scattering in the
thermal plasma. This scenario is refered to as
``thermal leptogenesis''.
However, the lightest $\nu_R$ can 
be produced only if their mass $M_1$ is less than the
reheat temperature $T_{RH}$ of the plasma
after inflation. In addition, the asymmetry is
proportional to $M_1$\cite{di2}, so there is a lower
bound on $M_1 $  to get a  large enough asymmetry.
This implies  $10^8$ GeV $ < M_1 < T_{RH}$.

The seesaw is an attractive minimal extension of
the SM that generates the observed small
$\nu$ masses. Three right-handed neutrinos, with large 
majorana masses $M_i$, are added to the Standard Model, along with a Yukawa
matrix for the neutrinos.  It is desirable to supersymmetrise
the seesaw, to address the hierarchy between the weak scale
and the $M_i$. In the SUSY seesaw,
$T_{RH}$ must be low enough to avoid over-producing gravitinos
\cite{gravitino1,gravitino2}---the canonical bound for gravity
mediated SUSY breaking 
is $T_{RH} \lappeq 10^9$ GeV. 
The aim of this paper is to identify the parameter
space where thermal leptogenesis can work, taking
 $M_1 \sim T_{RH} \sim 10^9$ GeV.

 The SUSY seesaw has more low-energy consequences than the non-SUSY
version. It
induces lepton flavour violating (LFV) entries in the
slepton mass matrix, which can lead
to radiative lepton decays\cite{Borzumati:1986qx}, 
such as $\mu \rightarrow e \gamma$,
at experimentally accessible rates. 
Eighteen  parameters  are required to define 
the neutrino and sneutrino mass matrices (in
the charged lepton mass eigenstate basis), which
is the same number as there
 are high scale inputs for  the seesaw model
\cite{Branco:gr}. It can be shown that
the SUSY seesaw can be parametrised with
the sneutrino and light neutrino  mass matrices\cite{Davidson:2001zk},
in a texture model independent way. That is, 
 the high-scale physical inputs of the
SUSY seesaw---the $\nu_R$ masses $M_i$ and Yukawa coupling
${\bf Y_{\nu}}$---can be ``reconstructed'' from
the neutrino and sneutrino mass matrices\footnote{This ``reconstruction''
would require universal soft masses at the GUT scale, and
improbable experimental accuracy at the weak scale, so is
in practice impossible.}. 
 The baryon asymmetry can
therefore be expressed as a function of weak scale
observables. In this paper we  identify the ranges
of experimentally measurable quantities which are
consistent with thermal leptogenesis.
This phenomenological analysis differs from previous
work  
\cite{review,real,other,phases,Falcone,Branco:2002kt,Endoh:2002wm,Plumacher:1997kc,Covi:1996wh,Barbieri:2000ma,Hirsch:2001dg,Branco:2002xf}
by making minimal assumptions
about the high scale theory: we assume the SUSY
seesaw and universal soft masses at the GUT scale. 
The usual approach is to assume
a GUT-scale texture that
generates the desired neutrino mass matrix, and discuss
leptogenesis---the aim here is to input the slepton mass matrix instead
of a texture. This
analysis should be consistent with all GUTS and texture choices
covered by these assumptions.

Section 2 includes notation,
and a review  of leptogenesis and our 
parametrisation of the seesaw model. Section 3 presents approximate
analytic expressions for the quantities on which
leptogenesis depends. The low energy signatures of the parameter space where
thermal leptogenesis works are discussed in section 4.
CP violation is briefly discussed in section 5.
 The results are discussed and summarised  in section \ref{sowhat}.

\section{Review}

The observed deficits in muon neutrinos from the atmosphere \cite{SK}
and in electron neutrinos from the sun\cite{Cleveland:1998nv,SNO,experiments} 
can be fit with small neutrino mass differences. The
recent KamLAND observation of a  $\bar{\nu}_e$ deficit from 
reactors confirms the neutrino mass explanation
of the solar neutrino puzzle\cite{Bahcall:2002ij}. The small
$\Delta m^2$ are consistent with three patterns of neutrino
mass: hierarchical ($\Delta m^2_{atm} = m_3^2$, 
$\Delta m^2_{sol} = m_2^2$), degenerate ($m_3 \simeq m_2 \simeq m_1 \gg
\Delta m^2_{atm}$) and quasi-Dirac ($m_3^2 \simeq m_2^2 \simeq
\Delta m^2_{atm}$, $\Delta m^2_{sol} = m_3^2 - m_2^2$). 
The leptogenesis scenario  considered in this paper, where the
$\nu_R$ are produced by scattering in the plasma, does
not work for degenerate $m_i$ \cite{di2} (see also
\cite{BdP} for a detailed discussion).  The quasi-Dirac
spectrum could be interesting, although it is possibly
disfavoured by supernova data \cite{SN}.
We assume the
$m_i$ are hierarchical, so the neutrino masses
are much smaller than the charged lepton and quark masses.
These small masses can be naturally understood in the seesaw model.

In subsection \ref{notn}, the seesaw is reviewed from the
top-down; introducing new physics at a high scale 
$M_X$, and seeing its low energy implications. This approach has been followed
by many model builders who construct  a natural
or symmetry-motivated structure of the high-scale 
mass and Yukawa matrices, and then
study its low energy consequences. See $e.g.$
\cite{Dreiner:1994ra} for early works that produce
neutrino mass matrices with small mixing angles, and 
\cite{Altarelli:gu} for more complete up-to-date references.
Lepton flavour violation due to the SUSY seesaw,
which could be observed in $\ell_j \rightarrow \ell_i \gamma$
\cite{Borzumati:1986qx,Hisano:1995cp}
or in slepton production and decay at colliders \cite{Arkani-Hamed:1996au}
has also been extensively studied from a top down
approach
(see $e.g$ citations of \cite{Borzumati:1986qx,Hisano:1995cp}).
Recent studies (for instance 
\cite{Lavignac:2002gf}) 
have considered the branching ratios
for $\ell_j \rightarrow \ell_i \gamma$ in models 
that induce the two observed large mixing angles among
the light leptons.\footnote{see also \cite{Casas:2001sr}
for a more phenomenological discussion of
$\mu \rightarrow e \gamma$}. 

Subsection \ref{lepto+ub}, is a ``top-down'' review  of leptogenesis, 
which is the   obvious approach 
\cite{review,real,other,phases,Falcone,Branco:2002kt,Plumacher:1997kc,Covi:1996wh,Barbieri:2000ma,Hirsch:2001dg,Branco:2002xf}
since the asymmetry is generated at high scales. See
\cite{review} for examples and models.
The translation between this approach and our bottom-up phenomenological
analysis is not obvious, so it is difficult to relate our work to
these papers. A phenomenological analysis of
leptogenesis in
non-SUSY (so no LFV) SO(10)
models was discussed in \cite{Branco:2002kt,Branco:2002xf}, with particular
attention to the low-energy CP violation.
A Yukawa-matrix independent analysis has
also been done in the case where there are only two
right-handed neutrinos\cite{Endoh:2002wm}.

\subsection{Notation and Numbers}
\label{notn}

We consider the supersymmetric see-saw for two reasons:
first, supersymmetry stabilizes the Higgs mass against the  quadratic
divergences that appear due to  heavy particles ($e.g.$ 
 the right-handed neutrinos). Secondly, 
the slepton masses enter  our bottom-up
parametrisation of the see-saw.

The leptonic part of the superpotential reads
\bea
\label{superp}
W_{lep}= {e_R^c}^T {\bf Y_e} L\cdot H_d 
+ {\nu_R^c}^T {\bf Y_\nu} L\cdot H_u 
- \frac{1}{2}{\nu_R^c}^T{\cal M}\nu_R^c , \eea
where $L_i$ and $e_{Ri}$ ($i=e, \mu, \tau$) are the left-handed 
lepton doublet and the right-handed charged-lepton singlet, 
respectively, and $H_d$ ($H_u$) is the hypercharge $-1/2$ ($+1/2$)
 Higgs doublet.
 ${\bf Y_e}$ and ${\bf Y_{\nu}}$ are the Yukawa couplings that 
give masses to the charged leptons and generate the neutrino Dirac mass, 
and $\cal M$ is a $3 \times 3$ Majorana mass matrix.
This is the minimal seesaw; additional terms are possible,
for instance in SO(10) models a small
triplet {\it vev} $ \langle T \rangle$ is probable\cite{Bajc:2002iw}, leading to a 
$\nu_L \langle T \rangle \nu_L$  mass term.

 We  work in the left-handed basis where 
the charged lepton mass matrix is diagonal, and
in a basis of
right-handed neutrinos where ${\cal M}$ is diagonal
\be
D_{\cal M} \equiv
{\mathrm diag}({ M}_1,{ M}_2,{ M}_3), 
\ee
with ${M}_i\geq 0$, and $M_1 < M_2 < M_3$.
 In this basis, the neutrino Yukawa matrix must
be  non-diagonal, but can always be diagonalized
by two unitary transformations:
\beq
\label{biunitary}
{\bf Y_\nu} = V_R^{\dagger} D_Y V_L,
\eeq
where $D_{\bf Y_\nu} \equiv
{\mathrm diag}({y}_1,{y}_2,{y}_3)$ and
$y_1 \ll y_2 \ll y_3$. Later in the paper, we
will assume that  $D_{\bf Y_\nu}$ is hierarchical,
with a steeper hierarchy than is in the light neutrino
mass matrix: $(y_1 /y_2)^2 \ll m_1 /m_2$.

It is natural to assume that the  overall scale of $\cal M$ 
 is much larger than the electroweak scale or any soft mass. 
Therefore, at low energies the right-handed neutrinos are decoupled and 
the corresponding effective Lagrangian contains a Majorana mass term
for the left-handed neutrinos:
$ \delta {\cal L}_{lep}= 
-\frac{1}{2}\nu^T{ m}_\nu \nu + {\rm h.c.},
$
%
%
%
with
\bea
\label{seesaw}
{m}_\nu= {\bf m_D}^T {\cal M}^{-1} {\bf m_D} =  {\bf Y_\nu}^T
{\cal M}^{-1} {\bf Y_\nu} \langle H_u^0\rangle^2.   \eea
We define the Higgs {\it vev} 
$\langle H_u^0\rangle^2=v_u^2=v^2 \sin^2\beta$, where   $v=174$
GeV. 
In the  basis where the charged-lepton Yukawa
matrix, $\bf{Y_e}$ and the gauge interactions are diagonal, the
$[m_\nu]$ matrix can be diagonalized by the MNS \cite{Maki:1962mu} matrix $U$ 
according to
\be
\label{Udiag}
U^T{[m_\nu] } U={\mathrm diag}(m_1,m_2,m_3)\equiv
D_{m_\nu}, \ee
where $U$ is a unitary matrix that relates  flavour to mass eigenstates
\bea  \pmatrix{\nu_e \cr \nu_\mu\cr \nu_\tau\cr}= U \pmatrix{\nu_1\cr
\nu_2\cr \nu_3\cr}\,,
\label{CKM}
\eea
and the $m_i$ can be chosen real and positive, and
ordered such that $m_1< m_2 < m_3$.
 Assuming hierarchical left-handed $\nu$ masses,
we take $m_3^2 = \atm = 2.7 \times 10^{-3} eV^2$
\cite{Hagiwara:fs}
 and  $m_2^2 = \sun= 7.0 \times 10^{-5} eV^2$ 
 \cite{B8}.
This corresponds to $m_3 = 5.2 \times 10^{-2}$ eV
($ 3.9- 6.3  \times 10^{-2}$  eV at  90\% C.L.),
and  $m_2 = 8.2 \times 10^{-3}$ eV
($7 - 15  \times 10^{-3}$ eV
at 3 $\sigma$).
$m_1$ is unknown, usually unimportant, and we take it to be
$m_2/10$. As we shall see, the  baryon 
asymmetry is weakly dependent on $\tan \beta$ in the parametrisation
we use, so we set $\sin \beta = 1$.

  $U$ can be written as
\bea U=V\cdot {\rm diag}(e^{-i\phi/2},e^{-i\phi'/2},1)\ \ ,
\label{UV}
\eea
where $\phi$ and $\phi'$ are CP violating phases,
 and $V$ has the form of the CKM matrix
\be \label{Vdef} V=\pmatrix{c_{13}c_{12} & c_{13}s_{12} & s_{13}e^{-i\delta}\cr
-c_{23}s_{12}-s_{23}s_{13}c_{12}e^{i\delta} & c_{23}c_{12}-s_{23}s_{13}s_{12}e^{i\delta} & s_{23}c_{13}\cr
s_{23}s_{12}-c_{23}s_{13}c_{12}e^{i\delta} & -s_{23}c_{12}-c_{23}s_{13}s_{12}e^{i\delta} &
c_{23}c_{13}\cr}.  \ee

The numerical values of the angles are
$.28 \leq \tan^2 \theta_{sol} \leq .91$ (3$\sigma $),
 with best fit point $ \tan^2 \theta_{sol} = .44$
\cite{B8}, so $\theta_{sol} = .41$.
We take  $\theta_{atm} = \pi/4$.
The CHOOZ angle  $\theta_{13}$ is experimentally constrained
$\sin  \theta_{13} \leq .2$ \cite{Apollonio:1999ae}.
Considerable effort and thought has gone into designing
experiments sensitive to smaller values of $\theta_{13}$.
J-PARC hopes to reach  $O(0.05)$ \cite{nufact},
and a neutrino factory could detect $\theta_{13}$
as small as $0.02 \rightarrow 0.001$\cite{nufact,Cervera:2000kp}.

We assume a simple gravity-mediated SUSY breaking scenario,
with universal soft masses at the  scale $M_X$. 
The sneutrino mass matrix (in the 
 charged lepton mass eigenstate
basis)  can be written in the leading log approximation as
\bea
\label{softafterRG} 
\left(m^2_{ \snu}\right)_{ij} & \simeq & 
 \left({\rm diagonal\,\, part}\right)
-\frac{3m_0^2 + A_0^2 }{8 \pi^2}
[ {\bf Y^{\dagger}_\nu} ]_{ik} [ {\bf Y_\nu }]_{kj} \ln \frac{M_X}{M_k}\ ,
\eea
where ``diagonal-part'' includes the tree level soft mass matrix, 
the radiative corrections from gauge and charged lepton 
Yukawa interactions, and the mass contributions from F- and D-terms.

The branching ratio for $\ell_j \rightarrow \ell_i \gamma$ can be estimated
\beq
\frac{BR(\ell_j \rightarrow \ell_i \gamma)}{BR(\ell_j \rightarrow \ell_i 
\bar{\nu}_i \nu_j)} \sim C \frac{\alpha^3}{G_F^2 m_{SUSY}^4}
 |y_k^2 \tilde{V}_{Lkj}^*\tilde{V}_{Lki}|^2  \tan^2\beta 
\simeq  10^{-7}  |y_3^2 \tilde{V}_{L3j}^*\tilde{V}_{L3i}|^2  \left( 
\frac{100GeV}{m_{SUSY}} \right)^4 \left( 
\frac{\tan \beta}{2} \right)^2
\label{BR}
\eeq
where $C \sim O(0.001 \div 0.01)$,  and $\tilde{V}_{L}$ diagonalises 
the second term of eqn (\ref{softafterRG}).
  More accurate formulae for the branching ratios can be
found in \cite{Hisano:1995cp}.
To further simplify these estimates, it would be
convenient  to assume that
$V_L = \tilde{V}_{L}$. That is,
 the lepton asymmetry will be  a function of
the angles of $V_L$, and it would be simplest to estimate
$\ell_j \rightarrow \ell_i \gamma$ using the angles
of $V_L$ for those of $\tilde{V}_{L}$.  
This a reasonable approximation
when   $\theta_{Lij} \gg \frac{y_i}{y_j} \theta_{Rij}$  $(i <j)$,
where $\theta_{Rij}$ ( $\theta_{Lij}$) is an angle of $V_R$($V_L$). 
For hierarchical Yukawa eigenvalues, this is likely to be true, even
if an angle $\theta_R$ in $V_R$ is large, because the usual
texture estimate for $\theta_{Lij}$ would be $\sqrt{y_i/y_j}$.
We assume  this  condition
is verified, so the principle
contribution to 
$[m^2_{ \snu}]_{ij}$ is
$\propto  y_k^2 {V}_{Lki}^* {V}_{Lkj}$.

Table \ref{tabmueg}  lists the current
upper limits on the $\mu \rightarrow e \gamma$,
 $\tau \rightarrow e \gamma$ and  $\tau \rightarrow \mu \gamma$
branching ratios, and the corresponding bounds on
$V_{L3j}$ that can be estimated from
eqn (\ref{BR}) \cite{Gabbiani:1989rb}.  
It also contains the hoped for sensitivity
of some anticipated rare decay searches. Colliders
could also be sensitive to flavour violating
slepton masses \cite{Arkani-Hamed:1996au}.

Leptogenesis will depend on angles of $V_L$. In the remainder
of the paper,  we will claim that ``leptogenesis predicts
an observable  $BR(\ell_j \rightarrow \ell_i \gamma)$'', if
the $V_L$ elements required exceeed the last column of
table \ref{tabmueg} (last three rows), with $y_3 = 1$.
It is clear that the branching ratios can be decreased, for
fixed $V_L$, by decreasing
$y_3$ and adjusting weak scale SUSY parameters. However,
if SUSY is discovered, these masses and mixing angles 
could in principle be measured at colliders, and
some information about the magnitude of $y_3$
could be available through the renormalisation group equations
\cite{Baer:2000hx}. This assumption of universal soft
masses at the scale $M_X$ will not be crucial  for our
conclusions.  Additional contributions to the off-diagonal
soft masses are unlikely to cancel the ones we discuss, so
the lower bounds we set on LFV branching ratios, from
requiring leptogenesis to work, should remain.

\begin{table}[hbt]
\begin{tabular}{|c|c|}
\hline
\hline
$BR(\mu \rightarrow e \gamma) < 1.2 \times 10^{-11}$ 
&$ y_3^2 V_{L32}^* V_{L31} < .006 $ \\
(PSI)  & \\
\hline
$BR(\tau \rightarrow e \gamma) < 2.7 \times 10^{-6}$ 
& $ y_3^2V_{L33}^* V_{L31} \lappeq 12$ 
\\
(CLEO)  & \\
\hline
$BR(\tau \rightarrow \mu \gamma) < 1.1 \times 10^{-6}$ 
& $ y_3^2 V_{L33}^* V_{L32} < 9 $
\\
(CLEO) &  \\
\hline
\hline
$BR(\mu \rightarrow e \gamma) \sim 10^{-14 \div 15}$ 
& $ y_3^2 V_{L32}^* V_{L31} \sim 3 \times 10^{-4} $\\
(PSI/nufact) &  \\
\hline
$BR(\tau \rightarrow e \gamma) \sim 10^{-9}$ 
& $y_3^2 V_{L33}^* V_{L31} \sim 0.2$ \\
(BABAR/BELLE)  & \\
\hline
$BR(\tau \rightarrow \mu \gamma) \sim 10^{-9 } $
&$y_3^2  V_{L33}^* V_{L32} \sim 0.2$ \\
(BABAR/BELLE/LHC)  & \\
\hline
\hline
\end{tabular}
\label{tabmueg}
\vskip 0.5cm
\caption{
Current limits \cite{Brooks:1999pu} 
 and hoped for sensitivities \cite{dreams} of some experiments.
The numerical bounds in the  right  column are multiplied by
$\left(\frac{m_{SUSY}}{100 ~{\rm GeV}} \right)^2 
\left(\frac{ 2}{\tan \beta} \right)$. If $\tan \beta$ is large,
these rare decays are sensitive to smaller angles in $V_L$ \cite{tomas}. }
\end{table}

Various CP violating phases
in  the neutrino and slepton mass matrices
could be measured in upcoming experiments.
However,  the experimental sensitivity to the phases
depends on the magnitude of unmeasured real parameters.
Anticipated $0 \nu \beta \beta$ experiments
may be sensitive to a maximal phase $\phi'$, for
the neutrino mass spectrum we consider.
The minimum value of the angle $\delta$ that could
be measured at a $\nu$ factory depends 
on $\Delta m^2_{32},\Delta m^2_{21}$ and $\theta_{13}$ 
(see $ e.g.$ \cite{Cervera:2000kp}),
so there is no  forseeable  clear upper bound.
The imaginary part of the product of off-diagonal slepton
masses $ \Im \{[m_{\snu}^2]_{12}[m_{\snu}^2]_{23}[m_{\snu}^2]_{31} \}
= \tilde{J} (\tilde{m}_2^2 - \tilde{m}_1^2) 
(\tilde{m}_3^2 - \tilde{m}_2^2) (\tilde{m}_1^2 - \tilde{m}_3^2) $
could be measured in slepton flavour oscillations
 down to $\tilde{J} = 10^{-3}$ \cite{Arkani-Hamed:1997km}.
 $\tilde{J}$ depends on the magnitude of the $[m_{\snu}^2]_{ij}$
as well as their phases, so all the phases
in the (s)lepton sector can be of order 1.

\subsection{Review of the parametrisation}
\label{revdi1}

 It was shown in \cite{Davidson:2001zk} that the seesaw can
be parametrised from the bottom-up, using
 the neutrino and sneutrino mass matrices. 
 See \cite{Ellis:2002fe} for  applications.
A similar phenomenological parametrisation of
the non-SUSY seesaw \cite{Broncano:2002rw} could be used,
if the scale $M$ of right-handed masses was low enough to
measure dimension six operators $\propto 1/M^2$.
We briefly review  \cite{Davidson:2001zk}  here. 

It is in principle possible to extract
the matrix 
\beq
P \equiv {\bf Y^{\dagger}_{\nu}}  {\bf Y_\nu} = V_L^{\dagger} D_Y^2 V_L
\label{step1}
\eeq 
from its contribution to the renormalisation group
running of the slepton mass matrix. 
This relies
critically on having universal soft masses at the
GUT scale, and on very  precise measurements of  sneutrino masses
and decays. It is therefore unrealistic\cite{Davidson:2001zk}.
However, since SUSY has not yet been discovered, 
$D_Y$ and $V_L$ can be used as  inputs
in a ``bottom-up'' parametrisation of the seesaw.

The aim is to determine  ${\bf Y_{\nu}}$ and ${\cal M}$ from 
$[m_\nu]$ and $P$.
$V_L$ and $D_Y$ can be determined from $P$, 
and used to strip the Yukawas off $[m_\nu]$:
\beq
\label{step2}
D_Y^{-1} V_L^* \frac{[m_\nu]}{v_u^2} V^{\dagger}_L D_Y^{-1} = 
V_R^*  D_{\cal M}^{-1} V^{\dagger}_R = {\cal M}^{-1} ,
\eeq
where the left hand side of this equation is known ($[m_\nu]$ is one of 
the inputs, and $V_L$ and $D_Y$ were obtained 
from eq. (\ref{step1})). Therefore, $V_R$ and
 $D_{\cal M}$ can also be determined.
This shows  that, working in the basis where the charged 
lepton Yukawa coupling, ${\bf Y_e}$,
the right-handed Majorana mass matrix, ${\cal M}$, and the gauge
interactions are all diagonal, it is possible 
to determine {\it uniquely}
 the  heavy Majorana mass matrix, ${\cal M}$, and the neutrino Yukawa 
coupling, ${\bf Y}_\nu = V_R^{\dagger}  D_Y V_L$,
starting  from $[m_\nu]$ and ${\bf Y}^{\dagger}_{\nu} {\bf Y_{\nu}}$.

\subsection{Leptogenesis, and the upper bound}
\label{lepto+ub}

The see-saw mechanism  provides a natural framework to generate the baryon
asymmetry of the Universe, defined as 
$\eta_B = (n_B - n_{\bar B})/s$, where $s$ is the entropy
density.
As was shown by Sakharov\cite{Sakharov:1967dj}, 
generating a baryon asymmetry requires baryon number violation, 
C and CP violation, and a deviation from thermal equilibrium.
These three conditions are fulfilled in the out-of-equilibrium
decay of the right-handed neutrinos and sneutrinos in the early 
Universe.
In the remainder of this paper,
``right-handed neutrinos'', and the shorthand notation
$\nu_R$, refer to  both  right-handed neutrinos and right-handed
sneutrinos.

In gravity mediated SUSY breaking scenarios,
these is an upper bound from gravitino production
on the reheat temperature  $T_{RH}$ of
the Universe after inflation.
The gravitino has a mass  $m_{3/2} \sim m_{SUSY}$  and only
gravitational interactions with SM particles, so it is very
weakly coupled, and long-lived. If a significant number of them decay
at or after Big Bang Nucleosynthesis, 
they could disrupt the predicted abundances
of light elements\cite{gravitino1,gravitino2}.
Gravitinos
can be created by various mechanisms
in the early Universe, 
such as scattering in the thermal plasma\cite{gravitino1,gravitino2}, 
or direct coupling to the inflaton (preheating) \cite{gravitino3,gravitino4}.
The latter is effective, but avoidable \cite{gravitino4}.
The number density of gravitinos produced in 
scattering increases with the plasma
temperature, so the bound on $n_{3/2}$ sets an
upper bound on the reheat temperature of the Universe after
inflation of
\beq
T_{RH} \lappeq  10^{9}  \rightarrow 10^{12} GeV 
\label{GRAV}
\eeq
(corresponding to $m_{3/2} \sim 100 $ GeV $ \rightarrow 10$ TeV
\cite{gravitino2}).  This bound assumes that the gravitino decays;
there are models where the gravitino is the LSP,
which allow $T_{RH} \lappeq   10^{11}$ GeV \cite{Berezinsky:kf}.

Let us briefly review the mechanism of generation of the BAU through
leptogenesis \cite{Fukugita:1986hr,review}.
At the end of inflation, a certain number density of right-handed
neutrinos, $n_{\nu_R}$, is somehow produced.
If these right-handed neutrinos  $\nu_{R_i}$  decay
 out of equilibrium,   a lepton asymmetry can be  created.
The subsequent ratio of the lepton excess to the entropy
density $s$ is given by
\beq
\eta_{L} = \frac{n_\ell - n_{\bar \ell}}{s} = 
 \sum_i \frac{n_{ \nu_{R_i} } }{s} ~ \epsilon_i ~ 
\tilde{d}_i.
\label{etaL}
\eeq
The CP-violating parameter
$\epsilon_i$
is determined by the particle physics model that gives
the masses and couplings of the $\nu_R$.
The value of $n_{\nu_R}/s$ depends on the
 mechanism to generate the right-handed neutrinos. We assume
the $\nu_{R_1}$ are generated by Yukawa scattering
in the thermal plasma, in which case
$n_{\nu_R}/s \lappeq  n^{eq}/s \simeq .2/g_*$,  where $n^{eq}$ is 
the  equilibrium number density of massless particles, and  $g_*
\simeq 230$  is the number of propagating states in the supersymmetric
plasma \footnote{in our conventions,
$n_{\nu_R} = (n_{\nu_R}+ n_{\bar{\nu}_R})/2$.
 The $.2$ is an approximation
to $g_*n^{eq}/s = \zeta(3) 135/(8 \pi^4)$ .}. 
 This also implies an upper bound on the $\nu_R$
mass: $M_1 \lappeq T_{RH}$. 
  Finally, $\tilde{d}_1$ is the
fraction of the produced asymmetry that survives
 after $\nu_R$ decay.  To ensure $\tilde{d}_1
\sim 1$, lepton number violating interactions (decays, inverse decays
and scatterings) must be out of equilibrium when the right-handed
neutrinos decay. In the case of
the lightest right-handed neutrino  $\nu_{R_1}$,
this corresponds approximately to
\beq
 K = \frac{\Gamma_{D_1}}{ 2H|_{T\simeq M_1}} < 1
\label{K}
\eeq
where $H$ is the Hubble parameter at the temperature
$T$, and $\Gamma_{D_1}$ the $\nu_{R_1}$ decay rate. 
There are two competing
requirements on the $\nu_R$ parameters---the
couplings must be large enough to produce a thermal
distribution, but small enough that the $\nu_R$
decay out of equilibrium.  Thermal leptogenesis
has been carefully studied in 
\cite{Plumacher:1997kc}\footnote{See \cite{Barbieri:2000ma}
for a detailed analysis of thermal leptogenesis
at higher temperatures, including the effects of
$\nu_{R2}$ and $\nu_{R3}$}. 
The numerical results of \cite{Plumacher:1997kc,review} 
suggest that $n_{\nu_R} \tilde{d}_1 /s < n_{\nu_R}^{eq}/s$: 
either $n_{\nu_R}$ does
not attain its equilibrium 
number density, or lepton number
violating interactions wash out a significant
fraction of the asymmetry as it is produced.
Defining an effective light neutrino ``mass''
\beq
\frac{\tilde{m}_1}{v_u^2} =  8 \pi
\frac{\Gamma_{D_1}}{ M_1^2}  = 
\frac{({\bf Y_{\nu} Y_{\nu}}^{\dagger})_{11}}{M_1}
\label{mtilde}
\eeq
$n_{\nu_R} \tilde{d}_1 /s \gappeq 10^{-4}$ is realised
for
\cite{BdP}
$5  \times 10^{-5} ~ {\rm eV}
\lappeq \tilde{m}_1 \lappeq 
 10^{-2}
$ eV.
The precise numerical bound on $\widetilde m_1$ 
 depends on $M_1$, and can be found
in \cite{Plumacher:1997kc}.

For 
$\tilde{m}_1 > 10^{-4}$ eV and $M_1 \sim 10^9$ GeV,
the dilution factor $d_1$  can be approximated
\cite{Asaka:2002zu,K+T}
\beq
\frac{n_{\nu_R} \tilde{d}_1}{ s}
\equiv d_1 \simeq
\frac{1}{6 g_*} \frac{1}{\sqrt{
K^2+ 1}}
\label{d1}
\eeq
with
$K 
\simeq 910\tilde{m}_1/eV  
$ from eqn (\ref{K}).
This is a
slight modification of the approximation, 
to ensure that it falls between the $M_1 = 10^8$ GeV and
$10^{10}$ GeV lines of \cite{review},  in the relevant range
$.001$ eV $ \lappeq \tilde{m}_1  \lappeq .1 $ eV. 
The exact numerical factor is important,
because it is difficult to get
a large enough asymmetry. Multiplying $d_1$ by a factor
of a few significantly increases the parameter space
where thermal leptogenesis can work. 
The approximation (\ref{d1})  neglects the decrease in $d_1$
at  $\tilde{m}_1 \lappeq 10^{-4}$ eV, which is due to underproduction
of $\nu_{R_i}$ in scattering. This is reasonable,
 because $\tilde{m}_1 \geq m_1$ \cite{di2}, 
and we take $m_1 = m_2/10$.

The last step is the transformation of the lepton asymmetry into a
baryon asymmetry by non-perturbative B+L violating (sphaleron)
processes
\cite{Kuzmin:1985mm}, giving 
\beq
\eta_{B} ={C} \eta_{B-L} =(3-9) \times 10^{-11},
\label{BAU}
\eeq
where $C = 8/23$  in the Minimal
Supersymmetric Standard Model.
Big Bang  Nucleosynthesis 
constrains $\eta_B$ to lie in the range of eqn (\ref{BAU}).
In a flat Universe,  the CMB determines
 $\eta_B \simeq (0.75  - 1.0) \times 10^{-10}$ \cite{boom}.
The wider BBN range is used in this paper, because
it is difficult to generate a large enough  $\eta_B$.

The CP asymmetry  can be approximated as
\bea
\epsilon_1 &\simeq&  -\frac{3}{8 \pi} \frac{1}{[{\bf Y_{\nu} Y_{\nu}}
^{\dagger}]_{11}} \sum_j {\rm {Im}} \left\{ [{\bf Y_{\nu}
Y_{\nu}^{\dagger}}]^2_{1j} \right\} \left( \frac{M_1}{M_j} \right) \\
&=& - \frac{3}{8 \pi}\frac{M_1}
{[{\bf Y_{\nu} Y_{\nu}}^{\dagger}]_{11}} {\rm {Im}} \left\{ [{\bf
Y_{\nu}} \frac{[m_\nu]^{\dagger}}{v_u^2} {\bf Y_{\nu}}^T]_{11} \right\}.
\label{eps1}
\eea
if the lepton asymmetry is  generated in the decay of the
lightest right-handed neutrino, and if the masses 
of the right-handed neutrinos are  hierarchical
\footnote{If the hierarchy in $Y_{\nu}$ is
similar to that of the quarks and charged leptons, then
a hierarchy in the $M_i$ is natural.}.
Were the asymmetry produced in the decay of
$\nu_{R_2}$ or $\nu_{R_3}$, it would depend on
a different combination of  couplings.

It is straightforward to show \cite{di2} that
if $\epsilon_1$ is written
\bea
|\epsilon_1|  = \frac{3}{8 \pi v_u^2}  M_1 {m_3} \delta_{HMY}
\label{bound}
\eea
then eqn (\ref{eps1}) implies the upper bound 
$\delta_{HMY} \leq 1$. The numerical results of
\cite{Hirsch:2001dg,Ellis:2002xg} agree with this
constraint.  
Using eqns (\ref{etaL}) and (\ref{BAU}),
this can be transformed into a lower bound
on $M_1$:
\bea
\label{boundM1}
M_1 \gsim  \frac{\eta_B}{C} 
\left[ \frac{n_{\nu_R}+ n_{\tilde{\nu}_R}}{s} ~ 
\frac{3}{8 \pi} \frac{m_3}{v_u^2} ~ \tilde{d}_1 \right]^{-1}
 = 
 10^{9} \left(\frac{\eta_B}{3 \times 10^{-11}} \right)
\left(\frac{.05 eV}{m_3} \right)
\left(\frac{4 \times 10^{-4}}{{d}_1 }\right)
{\rm ~GeV} . 
\eea
 Setting $m_3$ to its $90\% $ C.L. upper bound
0.063 eV, and
${d}_1$ to its maximum value $n^{eq}/s \simeq 45/(2 \pi^4 g_*)$,
implies $M_1 > 3 \times 10^8$ GeV.

This lower bound on $M_1$ comes very close to the gravitino bound
eqn (\ref{GRAV}) on the reheat temperature. Thermal
production of the $\nu_R$ requires $M_1 \lappeq T_{RH}$
so either $\epsilon$ is close to its upper bound, or
$M_1 ,T_{RH} > 10^9$ GeV, or thermal leptogenesis does not
generate the observed baryon asymmetry. We explore
the first option, and somewhat the second. The third
possibility, non-thermal $\nu_R$ production, has been
discussed by many authors (see {\it e.g.} references of
\cite{bdps}).

\section{Analytic approximations for $\delta_{HMY}$, $M_1$, $\tilde{m}_1$}
\label{secapprox}

At least three inputs are required to parametrise thermal 
leptogenesis \cite{Plumacher:1997kc,review,BdP}.
A possible choice would be the mass $M_1$  and
decay rate $\propto \tilde{m}_1$ of the $\nu_R$, and
the CP asymmetry $\epsilon_1$. 
However, $\epsilon \propto M_1$,  so we
use $M_1$,  $\tilde{m}_1$ and  $\delta_{HMY}$
(introduced by Hamaguchi, Murayama and Yanagida),
where  $\delta_{HMY}$ measures
how close $\epsilon$ comes to saturating its
upper bound. Note however, that $\delta_{HMY}$ is not a
CP phase.

This section contains simple analytic approximations
indicating the dependence of  leptogenesis parameters 
 on measurable  quantities, such as
neutrino masses and rare  LFV decays. We used this approximation,
with attention to the phases, in \cite{Davidson:2002em}.
A similar, somewhat simplified version was introduced
in \cite{Branco:2002kt}.

The inputs for the analytic approximation are:
\beq
 V_L, ~D_Y,~ U, ~[m_\nu]
\eeq
Two of the angles of $U$ are known, and the CHOOZ
angle is bounded above. The eigenvalues $y_i$  of
the neutrino Yukawa matrix are  unknown,
and realistically cannot  be determined from
the sneutrino mass matrix.
It seems reasonable to assume a hierarchy for the
$\{ y_i \}$, since we measure hierarchical Yukawas
for the quarks and charged leptons.
The  $y_i$ remain
as variables in the equations; we will discover
that only the smallest eigenvalue $y_1$ is
relevant, and can be ``traded'' for the mass
$M_1$ of
the lightest $\nu_R$, which is tightly
constrained.
$V_L$ contains three unknown angles, related
to the lepton flavour violating decays
$\ell_j \rightarrow \ell_i \gamma$.
 There are  three phases in both  U and $V_L$,
all are unknown, and 
assumed to be  chosen   to maximise the baryon asymmetry.

The lightest eigenvalue and 
corresponding eigenvector of
$\cal M$ are estimated in the first Appendix,
which also contains some simple (but illuminating)
3-d plots of leptogenesis parameters. 
The mass of the lightest $\nu_R$ is
\beq
|M_1| \simeq \frac{y_1^2v_u^2 }{|W^2_{1j} m_j| }.
\label{M1ap}
\eeq
where the matrix $W=V_L U$ is 
the rotation from the basis where the $\nu_L$ masses
are diagonal to the basis where the neutrino Yukawa  matrix
${\bf Y^{\dagger}_{\nu}} {\bf Y_\nu}$ is diagonal.
 There are three limiting values for $M_1$, 
corresponding to $M_1 \simeq y_1^2v_u^2/m_i$:
$M_1 \rightarrow  y_1^2v_u^2/m_3$ when $W_{13} \rightarrow 1$,
$M_1 \rightarrow  y_1^2v_u^2/m_1$ when $W_{13}, W_{12} \rightarrow 0$,
and   
$M_1 \rightarrow  y_1^2v_u^2/m_2$ when $W_{13}<m_2/m_3 $, 
$W_{12} \rightarrow 1$. This is easy to
see in  figure \ref{figM1}.

$M_1$ is the only quantity relevant for leptogenesis
which depends on $y_1$.
The latter is  effectively unmeasurable; it is constrained
by theoretical expectations, and by the requirement
that the analytic approximation be self-consistent.
Theoretically,  the eigenvalues of
$Y_\nu$ are expected to be hierarchical, and of order the quark
or lepton Yukawas, so 
figure \ref{figM1} is plotted 
with  $y_1 \sim 10^{-4}$. The approximations of
this section  are
consistent,  provided that the dropped $O(y_1^2, y_1^2/y_2^2)$ terms 
  are smaller than the  $O(m_1/m_3)$ terms
which are kept. This is the case for $y_1 \sim 10^{-4}$.
Since $y_1$ is unmeasurable and only weakly
constrained, it can be adjusted, as function of $m_i$ and $W_{1j}$,
to obtain a value for $M_1$ where leptogenesis
could work. In fact,
since $M_1 \propto y_1^2$ is tightly constrained,
the requirement $M_1 \sim 10^{9} $ GeV
 ``determines'' $y_1$.

The eigenvector  (\ref{eigenvec1}) can be used  
to evaluate  the $\nu_{R1}$ decay rate: eqn. (\ref{mtilde}) becomes
\beq
\tilde{m}_1 \simeq \frac{ \sum_k |W_{1k}^{2}| m_{_k}^2 }
{ |\sum_n W_{1n}^{2} m_{_n}| } 
\label{tildekappaap}
\eeq
 $\tilde{m}_1$ has various limits:  $\tilde{m}_1
\rightarrow m_3$ for $W_{13}$ large, 
$\tilde{m}_1
\rightarrow m_2$ for $W_{12}$ large and $W_{13} < m_2/m_3$ ,
and $\tilde{m}_1
\rightarrow m_1$ when 
$W \rightarrow  1$.
This is easy to see from the RHS of figure \ref{figM1}.
 In the $W \rightarrow  1$
limit, washout is minimised,  because  the dilution
factor $d_1 \propto 1/\tilde{m}_1$.

To saturate the upper bound (\ref{bound}),
$\delta_{HMY}$ needs to approach 1. 
Evaluating eq. (\ref{eps1}) with  the eigenvector(\ref{eigenvec1}),
gives 
\bea
\delta_{HMY} &=&
\frac{ {\rm Im} \left\{  \sum_{\ell,m} W_{1\ell}^{2} m_{\ell}^3
   W_{1m}^{*2} m_{m} \right\} }
 { m_3  |\sum_n W_{1n}^{2} m_{n}| (\sum_j |W_{1j}|^2  m_j^2)} 
\nonumber \\
&\simeq& \frac{  |W_{11}W_{12}|^{2} m_{1} m_{2}^3
+ |W_{11}W_{13}|^{2} m_{1}  m_{3}^3 
+  |W_{12}W_{13}|^{2} m_{2}  m_{3}^3 }
 { m_3 (\sum_n |W_{1n}|^2  m_n)  (\sum_j |W_{1j}|^2  m_j^2)}
\label{deltaapprox}
\eea
This paper is about  the relation
between real low energy observables (such
as $BR(\mu \rightarrow e \gamma)$) and the baryon asymmetry,
so scant attention will be paid to the phases in  $U$ and $ V_L$.
 For most of parameter space \footnote{
everywhere but when the three terms upstairs have equal
magnitude}, the phases can be chosen such that $\delta_{HMY}$
is larger than the second expression in eqn (\ref{deltaapprox}).

This second expression 
is plotted on the LHS in figure \ref{figeps3d}. 
 $\delta_{HMY}$ can approach 1 if the numerator
(upstairs) is dominated by  $m_3^3 m_2 $ or by
$m_3^3 m_1 $. This is
because of the $m_3$ in the denominator. 
If the $m_3^3 m_1 $ dominates
upstairs, then $\delta_{HMY}$ will approach 1 when $W_{1j}^2 m_j
\simeq W_{11}^2 m_1$ and  $|W_{1n}|^2 m_n^2
\simeq |W_{13}|^2 m_3^2$, or equivalently, when
\beq
\frac{m_1^2}{m_3^2}< W_{13}^2 < \frac{m_1}{m_3}
\, \,\, \, and  \, \,\, \,
W_{12}^2 < \frac{m_1}{m_2}, \frac{m_2^2}{m_3^2}
\label{aaa}
\eeq
This corresponds to the highest ridge in $\delta_{HMY}$ in
figure \ref{figeps3d}. Notice
that the position of the peak depends sensitively on  the lightest
neutrino mass $m_1$.

If  $m_3^3 m_2 $ dominates
upstairs, then $\delta_{HMY} \rightarrow 1$ 
 when 
\beq
W_{12}^2 \frac{m_2^2}{m_3^2 } < W_{13}^2 < W_{12}^2 
\frac{m_2}{m_3}
\, \,\, \, and  \, \,\, \,
W_{11}^2 < W_{12}^2\frac{m_2}{m_1} , W_{13}^2
\frac{m_3^2}{m_1^2}
\label{aab}
\eeq
This corresponds to the shoulder at slightly large $W_{13}$,
which is cut by $W_{12} \sim 1$ in the LH plot of figure 
\ref{figeps3d}. 

Finally, for $W_{13}$ very small, $\delta \rightarrow m_2/m_3 \sim0.1$ along
the ridge at $W_{12} \sim 0.1$. This corresponds
to the $m_2^3 m_1$ term dominating upstairs,
and arises when
\beq
W_{11}^2 \frac{m_1^2}{m_2^2 } < W_{12}^2 < \frac{m_1}{m_2 }W_{11}^2 
\, \,\, \, and  \, \,\, \,
W_{13}^2 < W_{11}^2\frac{m_1}{m_3} , W_{12}^2
\frac{m_2^2}{m_3^2}
\label{aac}
\eeq
Although $\delta_{HMY}$ does not reach its maximum value for
these parameters, the washout is small,
so the baryon asymmetry generated is only slightly
too small. As we will
see in figure \ref{figcont}, 
it is large enough if $M_1,T_{RH} \sim 10^{10}$ GeV
are allowed.

\section{When does thermal leptogenesis work?}
\label{when}

\begin{figure}[ht]
\includegraphics[width=7cm]{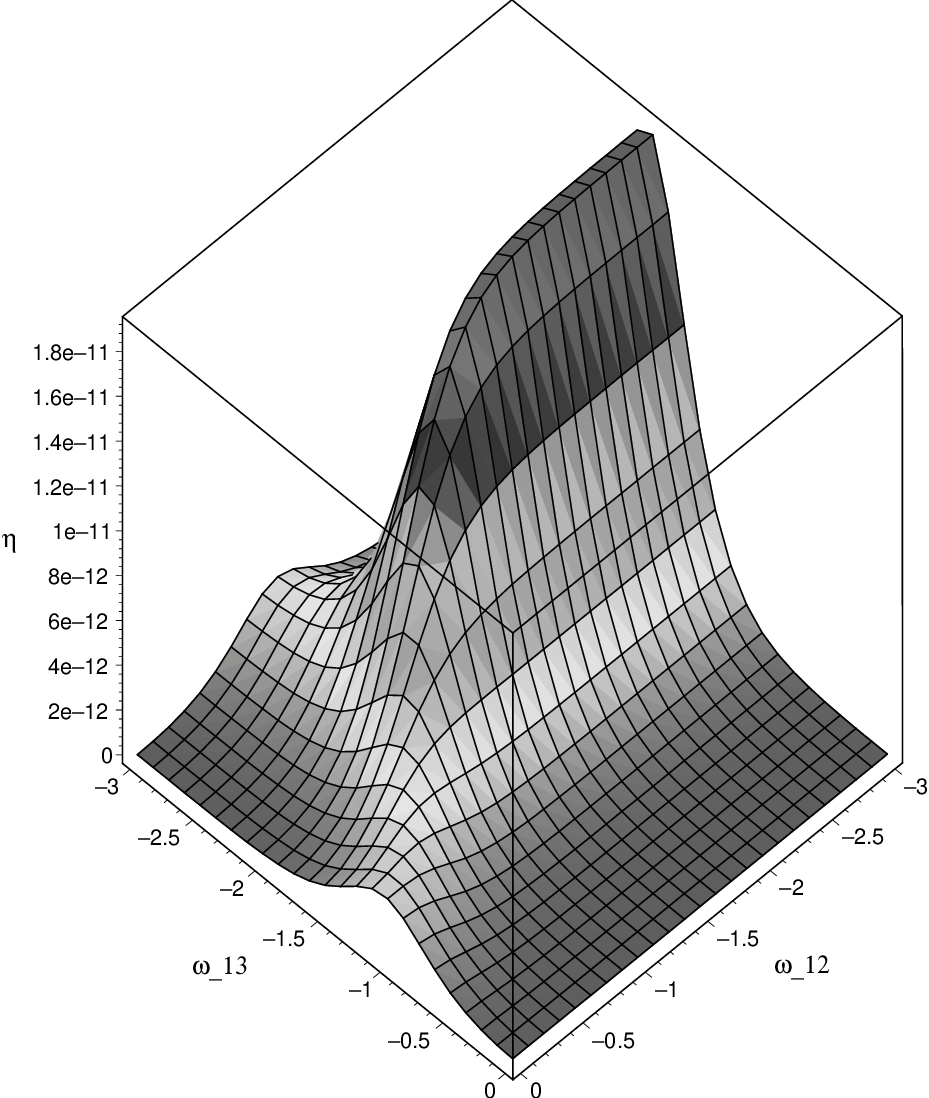}
\includegraphics[width=7cm]{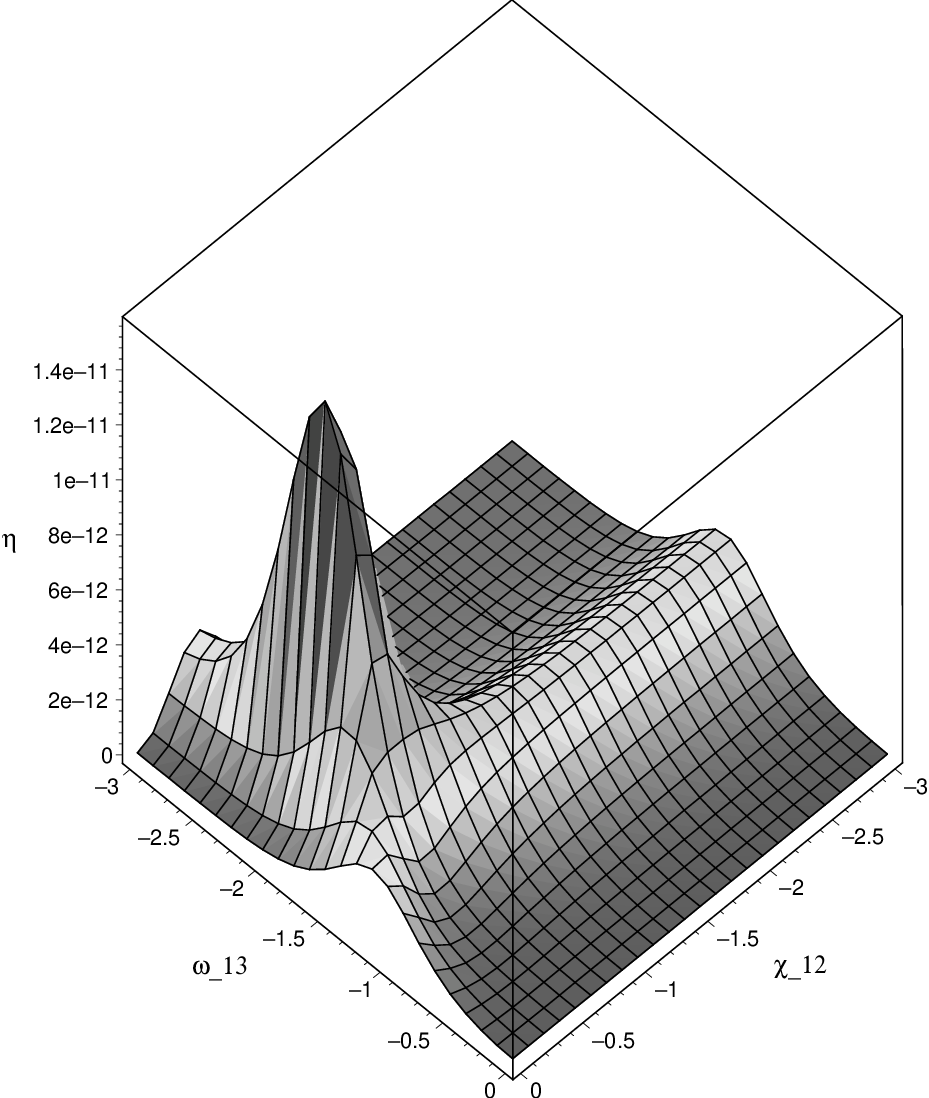}
\caption{3-d plot of $\eta_B= 8 d_1 \epsilon /23$, for
 central neutrino mass values, $m_1 = m_2/10$
 and $ M_1 = 10^9$ GeV.
On the left, $\eta_B$ is plotted  as a function of 
$\omega_{12} \simeq log [ W_{12} ]$ and $\omega_{13} \simeq log [ W_{13} ]$.
On the right, $\eta_B$ is plotted as a function of 
$\omega_{13}$ and 
$\chi_{12}$, defined such that $W_{12} = \cos \theta_{W13} \sin
( \theta_{sol} - 10^{\chi_{12}} \pi/2 )$. The RHS measure
on parameter space is more sensible, see the
discussion after eqn(\ref{W12}). }
\protect\label{figetaB}
\end{figure}

The baryon asymmetry can be approximated as
\beq
\eta_B \simeq \frac{8 d_1}{23} 
\frac{3}{8 \pi v_u^2} M_1 m_3 \delta_{HMY}
\label{etaBap}
\eeq
by combining eqns (\ref{etaL}), (\ref{BAU}), and    (\ref{bound}).
 This is plotted in figure
\ref{figetaB}, which  suggests that $\eta_B$ $can$ be large
enough. 

The issue is whether a large enough
asymmetry can be generated, so
 the observational upper limit on
$\eta_B$ is unimportant. Also, the asymmetry calculated
here is the upper bound corresponding to
maximal CP violation, so it can be
reduced by taking smaller phases.
We use the one-$\sigma$ observational lower bound
on $\eta_B$ from nucleosynthesis:
 $ \eta_B \gappeq  3 \times 10^{-11}$.
To obtain a large enough baryon asymmetry by
thermal leptogenesis,
the parameters $M_1$, $\delta_{HMY}$, and  $d_1$
 must occupy narrow ranges. 
The washout effects
are minimised when the $\nu_R$ decay rate is small, which corresponds to
$W \rightarrow 1$. In this case, 
$n_{\nu_R} \tilde{d}_1/s = d_1 < 10^{-3}$
which implies the lower bound 
 $\epsilon \gappeq 10^{-7}$. 
(If   $\epsilon  \gappeq 10^{-6}$ can be obtained, then
$  d_1 \sim  10^{-4} $ is large enough.)
Eqn (\ref{bound}) implies a lower bound
on $M_1$  to get $\epsilon_1$ large enough. In addition,
$M_1 \lappeq T_{RH}$ is required for
thermal production; the canonical SUSY gravitino bound
is  $T_{RH} \lsim 10^9$ GeV, so 
$
5 \epsilon  \times 10^{15} {\rm GeV} \lappeq M_1 \lappeq T_{RH}
$. Since  $\epsilon \simeq 10^{-7}$ is required,
for   $M_1 \lsim 10^9$ GeV
one must have  $\delta_{HMY} \rightarrow 1$.
The parameter space of choice can be
summarised as
\bea
few \times 10^{8} GeV & \lappeq M_1 \lappeq &  few \times 10^{9} GeV  
\nonumber \\
  d_1 & \rightarrow &  \frac{45}{2 \pi^4 g_* } ~~  , g_* = 230 
\label{desire} \\
 \delta_{HMY} & \rightarrow & 1
\nonumber
\eea 
As can be seen from figures \ref{figM1} --- \ref{figeps3d},
it is difficult to simultaneously satisfy these
conditions. $M_1$ and $d_1$ increase as $W_{13}, W_{12} \rightarrow 0$,
but $\delta$ decreases.

The analytic approximations of the previous section
show that the baryon asymmetry depends on $U$ and the first row
of $V_L$ (via $W_{1j}$), on the light neutrino masses
$m_i$,  on the lightest neutrino Yukawa $y_1$,
and on phases.
These real parameters are known, or
could  be experimentally  constrained
in the next 20 years---with the exception of
$m_1$, $ y_1$, and $V_{L12}$.
 For
the purposes of this paper, 
 $V_{L12}$ is included with the measurable angles,
and  $ ~y_1$ is $determined$ as a function of  $m_1$,
by requiring that $M_1$  be in the range (\ref{desire}) 
where leptogenesis could work.
The baryon asymmetry then becomes independent
of $y_1$. Some subtle dependence on
$m_1$ remains:
the area and location of the high ridge in
figure \ref{figetaB} depend on $m_1$, but
the baryon asymmetry and low energy
footprints do not.
This is  discussed in the Appendix about $m_1$.

Notice also that it could be expected to have a similar
hierarchy in the neutrino Yukawas as in the other fermions,
in which case $y_1  \sim h_u, h_e$ or $h_d$.  This gives
\beq
M_1 \sim \left( \frac{y_1}{h_u} \right)^2 \left( \frac{m_2}{W_{1j}^2 m_j} \right)  3 
\times 10^6 ~ {\rm GeV}
\eeq
where $W_{1j}^2 m_j$ is usually of order  $m_2$.
If $y_1 \simeq h_u$, then $\eta_B$ is too small
over most of  parameter space. This was found
in some models by \cite{Falcone}. The
baryon asymmetry  can be large enough,
for  $y_1 \simeq h_u$, in the small
area of parameter space where 
$W_{1j}^2 m_j \simeq m_1 \lappeq m_2/100$.
This is  in the $m_1$ Appendix too.

The baryon asymmetry depends weakly on $\tan \beta$,
 when  $M_1$  is taken as an input, and  
$d_1$ is approximated as 
$\propto 1/\tilde{m}_1 \propto \sin^2 \beta$. The $m_i$
are experimentally measured, and therefore independent
of  $ \sin^2 \beta$, so it is clear from
eqn (\ref{etaBap}) that the $\sin \beta$ dependence
arises entirely from  $\tilde{m}_1$.
 If   instead
$M_1 = (y_1^2v_u^2)/|W_{1j}^2 m_j|$,
then $\eta_B \propto \sin^4 \beta$.
In both cases,  larger  $\sin \beta$ is marginally favoured.

The parameters $W_{12}$ and  $W_{13}$ are convenient,
because they summarise the unknown mixing angles and phases.
The  physically relevant quantities for leptogenesis
($M_1, \epsilon,$...) can be plotted as a function of the two
real unknowns $|W_{12}|$ and  $|W_{13}|$.  However, 
the $W_{1j}$  are not observable---the matrix $W$ is related
to the more physical matrices $V_L$ and $U$ by $W = V_L U$.
Recall that $V_L$ rotates from the basis
where  the neutrino Yukawa matrix
$Y_\nu$ is diagonal to the basis  where $Y_e$ is diagonal,
and $U$ rotates from the basis where $[m_\nu]$ is
diagonal to the basis where $Y_e$ is diagonal.
$W$ can be written
\bea
W_{13}&  = &  V_{L11} \sin \theta_{13}e^{-i\delta}  +
 V_{L12}/\sqrt{2} + V_{L12}/\sqrt{2}
\label{W13} \\
W_{12}&  =&   V_{L11} 
\sin \theta_{sol}
 + 
V_{L12} ( \cos \theta_{sol} -   \sin \theta_{sol} 
\sin \theta_{13} e^{i \delta} )/\sqrt{2}  \nonumber \\
& & 
- V_{L13}( \cos \theta_{sol} +   \sin \theta_{sol} 
\sin \theta_{13} e^{i \delta} )/\sqrt{2}  
\label{W12}
\eea
where $\theta_{12} = \theta_{sol}$ and 
 $\theta_{23} = \pi/4$ in the MNS matrix.

From a model building perspective\cite{Altarelli:gu}, there are two natural
limits for $W$. The most popular is for  the large leptonic  mixing
angles to come from the seesaw
structure of the light neutrino mass matrix. In this case,
$V_L \sim 1 $ can easily arise, so $W \sim U$. This is similar to
the quark sector, where the CKM matrix (the analogue of $V_L$)
has small angles.  An example
of this is  texture models where the large
atmospheric mixing angle is due to a  $\nu_R$ mass eigenstate having
approximately equal Yukawa couplings to $\nu_\mu$ and $\nu_\tau$
\cite{Barbieri:1998jc}.
Alternatively, the  electron Yukawa matrix $Y_e$ could
be ``the odd man out'' ; it could have large
off-diagonal elements in a basis where the neutrino  mass matrix $[m_\nu]$
and $Y_\nu$ are simultaneously almost diagonal\cite{Babu:1995hr}.
In this case $V_L \sim U^\dagger$ and $W \sim 1$.
These two cases are discussed in the following two
subsections. Figure \ref{figetaB} suggests that thermal
leptogenesis can work for  $W \sim 1$. 
As we shall see, a
 large enough asymmetry is also possible 
in the $V_L \rightarrow 1 $ limit,
if a slightly larger $T_{RH}$ is allowed.

The plots are functions of $\log W_{13}$ and $\log W_{12}$,
rather than, $e.g. W_{12}$ and $W_{13}$.
It is sensible to use
logarithmic measure on unknown physical parameters\footnote{ this
choice of measure is neither unique nor universally agreed on}
because it is equally probable for mixing
angles to have any order of magnitude between
$e.g. 10^{-3}$ and 1. However, $W_{1j}$ are not physical
parameters, so this reasoning does not apply to them. 
Specifically, values of $W_{12} \lappeq \sin \theta_{sol}$
arise in the presumably small area of parameter space
where $V_L \simeq U^\dagger$. 
This reasoning does apply to
the CHOOZ angle 
and the unknown angles of $V_L$, 
but  we prefer to plot $\eta_B$ as a function of two
unknowns, rather than four.
So a more appropriate measure on $W_{12}$ might be
logarithmic in the difference
away from $U_{12} \simeq \sin \theta_{sol}$.  Therefore,
 on the RHS of figure \ref{figetaB} is plotted  the same function as
on the LHS, but as a function of $\omega_{13} \simeq$
 log($W_{13}$),  and $\chi_{12}$, the latter
defined such that 
\beq
W_{12} = \sin ( \theta_{sol} - 
10^{\chi_{12}} \pi/2 )
.
\label{defchi}
\eeq
$\chi_{12} \simeq \log [W_{12} -  U_{12}]$
 is an approximation to the log of the unknown
angles of $V_L$ (see eqn (\ref{W12})). So the RHS plot tells
us the same information as its twin on
the left: the asymmetry is largest
if a large angle in $V_L$
cancels the large solar angle in the MNS matrix $U$.

A final technical comment: $W\sim 1$ and $V_L \sim 1$ mean
the 12 and 13 matrix elements 
are small $ \lsim .1$. $W \sim 1 $  means
that $W$ maximises $\eta_B$, so $W_{12} \lsim .1$ and
 $.01 \lsim W_{13} \lsim .1$.  $V_L \sim 1 $ 
includes both the possibilities that the angles
of $V_L$ are smaller, or larger, than the CHOOZ angle.

Section \ref{Wsim1}  studies the phenomenological
consequences of sitting in the region
where thermal leptogenesis works easily,  which corresponds
approximately to 
$W_{12} \lsim .1$, $.01 \lsim W_{13} \lsim .1$. 
Then in section \ref{VL=1},  some of the parameters
which are fixed in figure \ref{figetaB} are varied, 
so  a large enough baryon asymmetry
can be generated for $V_L = 1$. 
The parameter space between these two limits 
is discussed in section \ref{VLneq1}.

\subsection{$W \sim 1$}
\label{Wsim1}

The parameter space  where $\eta_B$ is largest
in figure \ref{figetaB}
corresponds roughly to 
\beq
.01 \lappeq W_{13}  \lappeq .1
~~~~W_{12}  \lappeq .1 ~~~.
\label{W1jap}
\eeq
This can be understood from the analytic
approximation (\ref{etaBap}). We fix $M_1 \simeq 10^{9}$ GeV,
so $\eta_B \propto d_1 \delta_{HMY}$. The  factor
$d_1$ is largest when the $\nu_R$ decay rate 
$\Gamma \propto \tilde{m}_1$
 is smallest,
so more of the asymmetry survives when $W \rightarrow 1$
(see the expression \ref{tildekappaap}).
$\delta_{HMY}$ parametrises how close $\epsilon$
can come to its upper bound (\ref{bound}).
For $m_1 \sim m_2/10$, the values of $W_{12}$, $W_{13}$ 
 where $\delta_{HMY}$ is maximised 
(eqn \ref{aaa}) correspond to eqn (\ref{W1jap}).
$\eta_B$  is maximal at smaller $W_{13}$ than $\delta_{HMY}$,
as can be seen by comparing figures \ref{figetaB} and \ref{figM1}.
This is due to the lepton number washout encoded in $d_1$,
which is faster at larger  $W_{13}$.

To obtain $W_{12}$ and $W_{13}$ in this region, $V_L$
must have the form
\beq
V_L = R_{23} [ U_\approx]^\dagger
\label{W1VL}
\eeq
where $R_{23}$ is an unspecified complex rotation in the 23
plane (written in  the form of eqn (\ref{Vdef}) with $\theta_{12} = 
\theta_{13} = 0$,  $S \equiv \sin \theta_{23},$
$C \equiv \cos \theta_{23}$, and taking a
23 phase  $\alpha$), and $[U_\approx]$ is a matrix
whose angles are roughly those of the MNS matrix $ \pm .1$.
The unknown $R_{23}$ appears because leptogenesis only depends
on the first row of $W$.

It is interesting to study the implications for
$\ell_j \rightarrow \ell_i \gamma$ of eqn (\ref{W1VL}).
Taking  $[U_\approx] = U$
\bea
V_{L31}& =& S e^{i(\alpha + \phi'/2) }
 c_{13}s_{12} + C  s_{13}e^{i\delta} \nonumber \\
& \simeq & S e^{i(\alpha + \phi'/2)}s_{sol} + C s_{13}e^{i\delta} \nonumber\\
V_{L32}& =&  S e^{i(\alpha + \phi'/2)}( c_{23}c_{12}-s_{23}s_{13}s_{12}e^{i\delta})
+ C  s_{23}c_{13} \nonumber \\
& \simeq & S e^{i(\alpha + \phi'/2)} c_{sol}/\sqrt{2} + C/\sqrt{2} \nonumber
\eea
For generic values of $S$,  this implies $V_{L32} \sim 1$,
so an  experimentally accessible  $\tau \rightarrow \mu \gamma$
branching ratio. 
If,
on the other hand, 
$S$ is tuned  to make $V_{L32} \rightarrow 0$,
then $BR(\tau \rightarrow \mu \gamma)$
would be  unobservable.  However, in this case
 $V_{L31} \sim - \sin \theta_{sol}/\sqrt{1 + \cos^2 \theta_{sol}}
\simeq 1/\sqrt{3}$, so $\tau \rightarrow e \gamma$ should be observable. 
One can conclude that if leptogenesis takes place in
the $W \sim 1$ peak of figure \ref{figetaB}, then
one or both of  $\tau \rightarrow \mu \gamma$
and $\tau \rightarrow e \gamma$ should 
have  a branching ratio $\gappeq 10^{-9}$ (according 
to the leading log approximation of the introduction).
Similarly,  $BR(\mu \rightarrow e \gamma)$ should be
$\gappeq 10^{-14}$ if $S $ or $s_{13} \gappeq 10^{-3}$.

\subsection{$V_L = 1$}
\label{VL=1}

It is barely possible to get a large enough baryon
 asymmetry when the angles in $V_L$ are small,
although this is not evident from figure \ref{figetaB}.
In the limit $V_L \rightarrow 1$, the matrix $W \rightarrow U$,
so $W_{12} \simeq \sin \theta_{sol}$ and 
 $W_{13} \simeq \sin \theta_{13}$.
In figure \ref{figetaB}, $\eta_B$ is at least a factor of
6 to small at $\log W_{12} \sim -0.5$ (equivalently,
$\chi_{12}$ small),
 but there is a bump 
at $W_{13} \sim .1$.  In this section, 
 $\eta_B$ at $V_L = 1$ is increased  by varying
$m_3, m_2$ and $M_1$;
 $V_L $ close to the identity is discussed
in the following subsection.

  If $V_L =1$,  eqns  (\ref{M1ap}) and (\ref{tildekappaap}) give
\bea
M_1 &  =& \frac{y_1^2 v_u^2 }{m_2 s_{12}^2} \nonumber \\
\tilde{m}_1 & =& m_2 \nonumber 
\label{MmtV1}
\eea
To maximise the asymmetry,  $M_1$ is taken  to be $ f \times 10^9 $ GeV,
where $f$ is a few 
(this determines $y_1^2  =  M_1 m_2 s_{12}^2/v_u^2 \simeq 7.2 f \times
10^{-8}$ ). In  figure 10  of \cite{review},  the lepton
asymmetry is plotted as a function of $\tilde{m}_1$ for
various values of $M_1$ and $\epsilon_1 = 10^{-6}$.
This plot shows  that for $\epsilon \simeq 10^{-6}$, 
a large enough asymmetry can be generated
when  $\tilde{m}_1 \simeq m_2$. 
Note that the washout
effects are correctly included in this plot
of \cite{review}, so this result does not depend on the analytic
approximation of eqn (\ref{d1}). Also, the asymmetry
{\it increases} as $\tilde{m}_1$ decreases, so
small values of the solar mass 
are prefered.

\begin{figure}[ht]
\includegraphics[width=7cm]{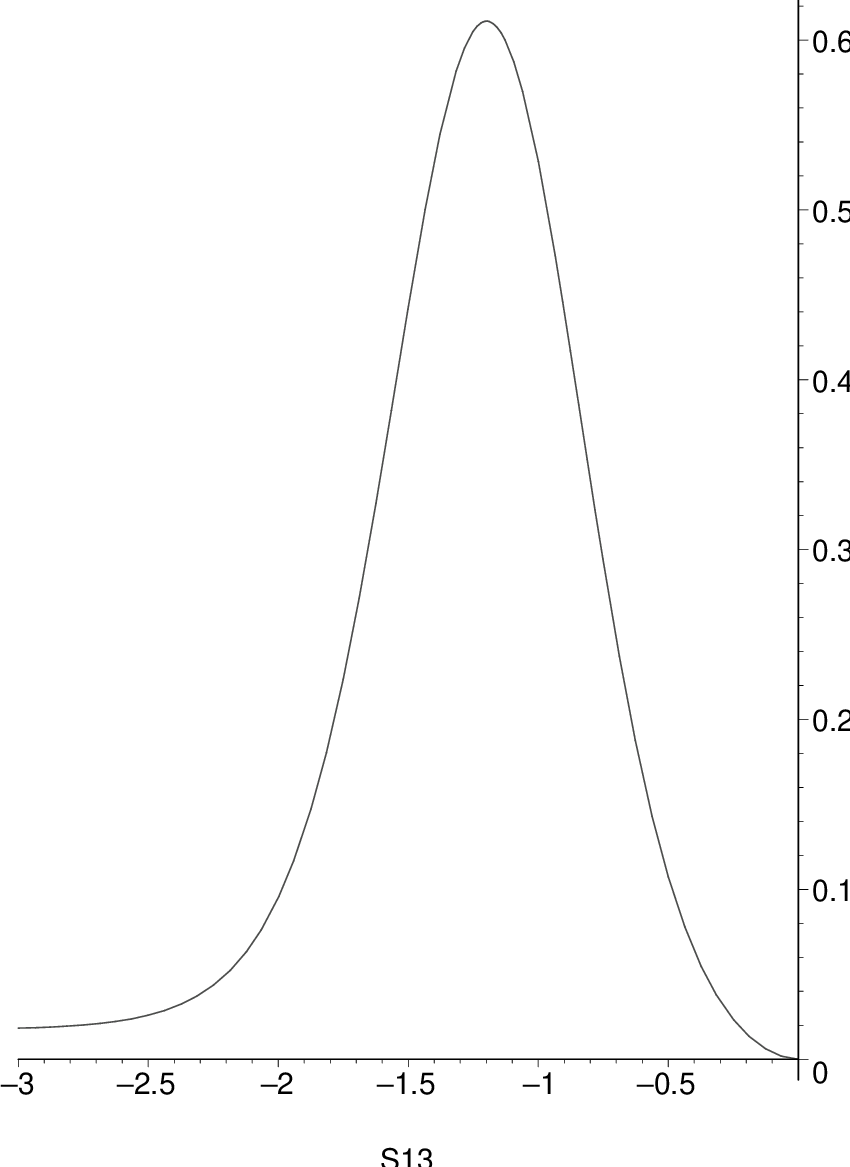}
\includegraphics[width=7cm]{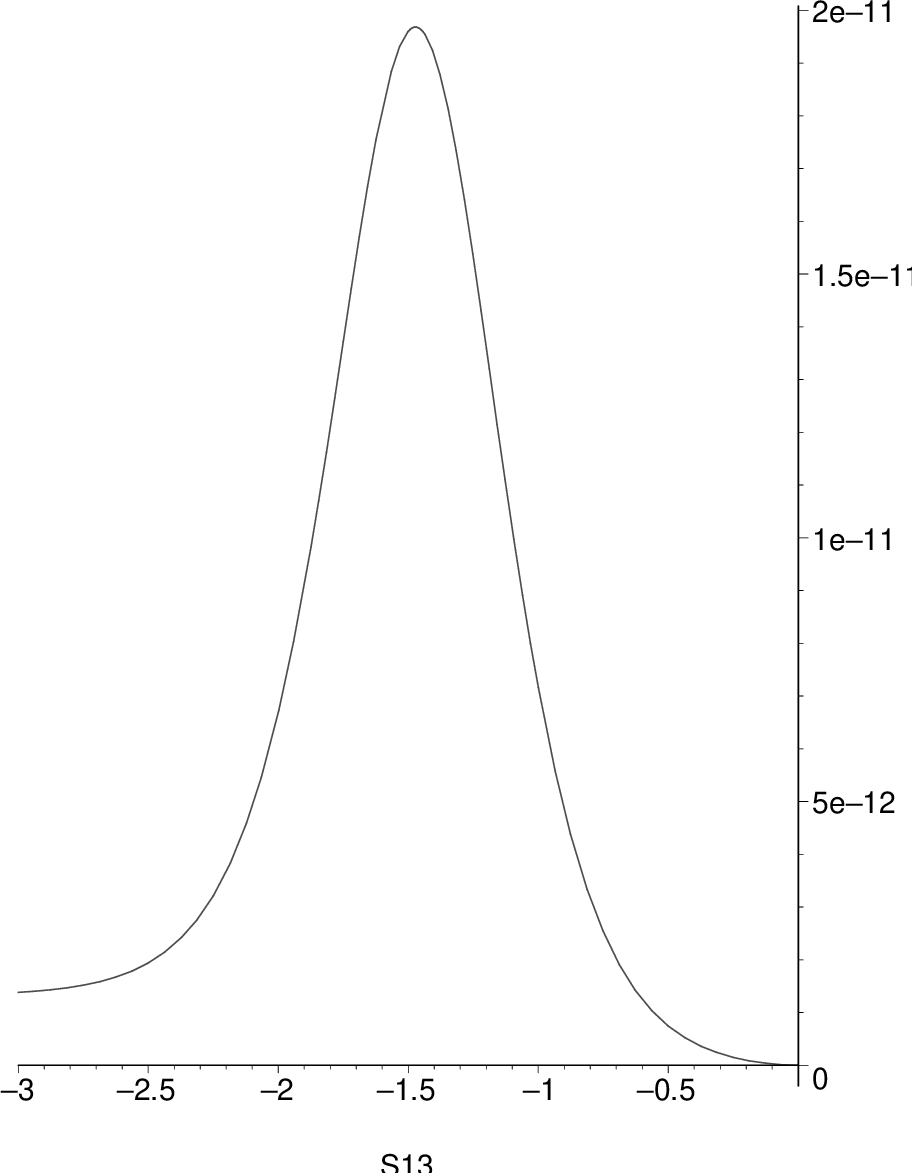}
\caption{
On the LHS $\delta_{HMY}$, and 
on the RHS the analytic approximation
to $\eta_B$, as a function of 
$ S{13} = log [ \sin \theta_{13} ]$.
$\delta_{HMY} \sim 1$
is required to get a large enough asymmetry,
see the discussion in section \ref{VL=1}.
The remaining parameters are  $ \tan^2 \theta_{sol} = 0.44 $,
 $m_2  = 7 \times10^{-3}$ eV, 
$m_3  =  6.3 \times 10^{-2}$ eV,
and for the $\eta_B$ plot,
$V_L = 1$ and $M_1 =
4 \times 10^{9}$ GeV.  }
\protect\label{figsol}
\end{figure}

The upper bound of eqn (\ref{bound}) implies that
for $\epsilon = 10^{-6}$, $f m_3 \delta_{HMY} =
0.25 $eV. From the experimentally allowed
range of $m_3$ given after eqn  (\ref{CKM}), we see that
$M_1 \sim 3 \times 10^{9}$ GeV is required, assuming
$\delta \sim 1$ is also possible.  For $V_L = 1$,
\beq
\delta_{HMY} \simeq \frac{ s_{13}^2 m_3^3 s_{12}^2 m_2 \sin (\phi'
- 2 \delta)}
{m_3(s_{12}^2 m_2^2 + s_{13}^2 m_3^2) s_{12}^2 m_2}
\label{dV1}
\eeq
which approaches 1 when $ s_{13} \simeq s_{12} m_2/m_3$.
In the RH plot of figure (\ref{figetaB}), $V_L = 1$
corresponds to $W_{13} = \sin \theta_{13}$ and $\chi_{12}
\rightarrow - \infty$.
 In figure \ref{figsol},  $\delta_{HMY}$ 
is plotted as a function of $\theta_{13}$ on
the LHS; 
  the analytic approximation to $\eta_B$ 
(eqn (\ref{etaBap})) is plotted on
the RHS.
So thermal leptogenesis
``works'' at $V_L = 1$, for $M_1 \sim 6 \times 10^{9}$ GeV.

Phrased another way:
for arbitrarily small  $\ell_j \rightarrow \ell_i \gamma$
branching ratios, requiring thermal leptogenesis to  work
{\it predicts}  the CHOOZ angle $\theta_{13}$.  If  $m_3$ 
is  taken at its $90 \% C.L.$ upper bound, 
and  $M_1 \sim 6 \times  10^{9}$ GeV, then
$\eta_B \sim 3 \times 10^{-11} $ can be obtained.
This requires a CHOOZ angle of
$\theta_{13} \sim 4 \times 10^{-2}$,
and phases which  satisfy
 $2\delta - \phi' = \pi/2$.

\subsection{$V_L$ from $1$ to $U^\dagger$}
\label{VLneq1}

It is clear from the RHS of figure \ref{figetaB}, that if a large enough
asymmetry can be generated at $V_L = 1$ ($\chi_{12} \sim -3$),
 then enough baryons
can be generated along the  
ridge leading to the  peak, and  also along the ``other'' ridge
at $W_{12} \sim m_1/m_2$. To sit on this  second ridge
requires $W \sim 1$, so has the same experimental signatures
as discussed in  section \ref{Wsim1}.
This section  is about  the 
$W_{13} \sim m_2/m_3$ ridge stretching from $V_L \sim 1$ to
the $W \sim 1$ peak.

Starting  from the small $\chi_{12}$, flat section of the ridge
and moving  towards the peak
corresponds to allowing small matrix elements $V_{L12}, V_{L13}
\lsim .1$. 
In this limit,
\bea
W_{13}
 & \simeq &  \sin \theta_{13} + V_{L12}/\sqrt{2} + V_{L13}/\sqrt{2} 
\nonumber
\label{W1jap2}
\eea
and $W_{13} \sim 0.04$ is required to get a
large enough asymmetry. Unfortunately, 
$W_{13} \sim 0.04$ determines a sum of three
unknowns: $\theta_{13}
\gsim .04$ could be observed, $V_{L13} \sim 0.04$ induces a potentially
observable $\tau \rightarrow e \gamma$ signal, but
$V_{L12} $ has no observable consequences.  For $V_L \sim 1$,
the $V_{L12}$ contribution to $m_{\snu}^2$ (eqn (\ref{softafterRG}))
is suppressed by $y_2^2 \sim 10^{-4}$.

 \begin{figure}[ht]
\includegraphics[width=10cm]{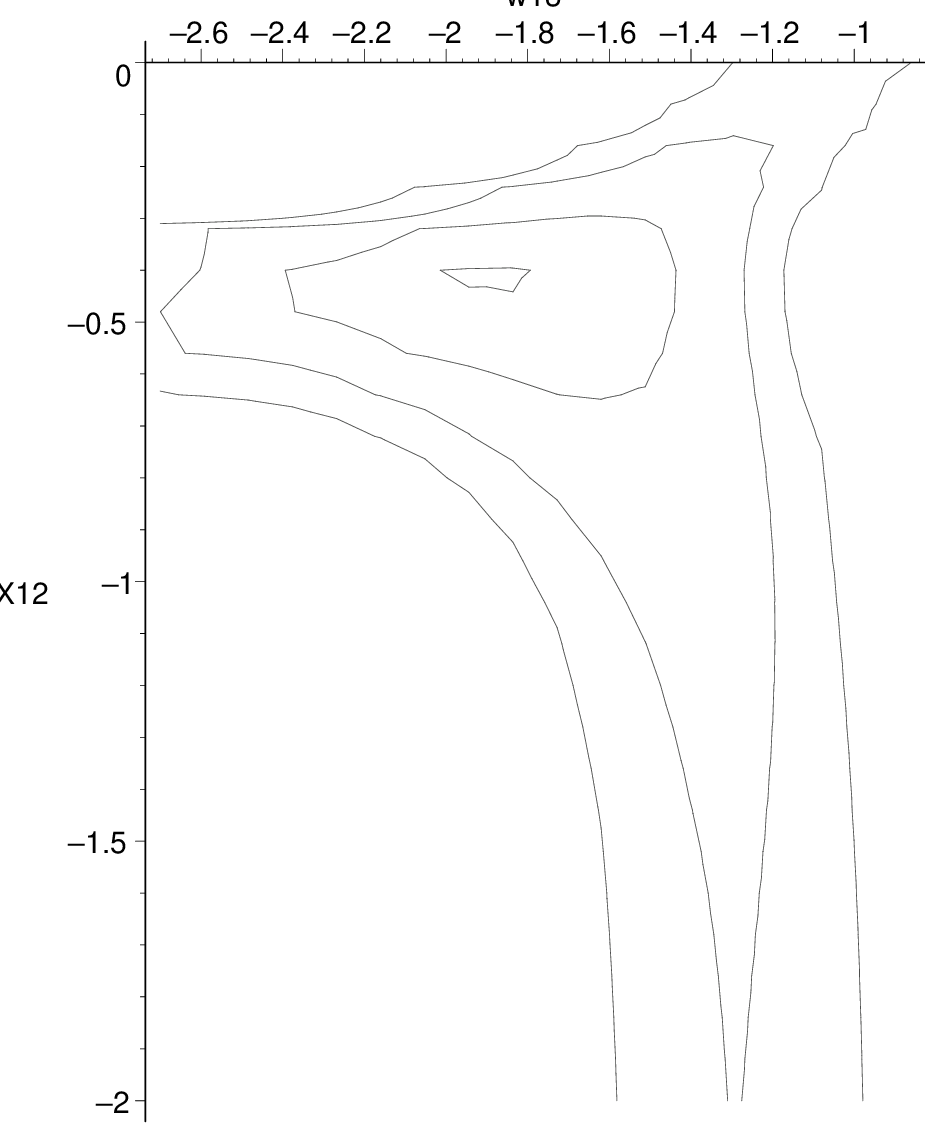}
\caption{Contour plot  of $\eta_B$, 
as a function of $\omega_{13} \simeq \log [W_{13}]$  and 
$\chi_{12} \simeq \log [ V_{L12} + V_{L13}]$.  
 The contours enclose the area when 
$\eta_B > 2 \times 10^{-11}$, for $M_1 =f \times  10^9 $ GeV,
central values of $m_3$ and $m_2$,  and $m_1 = m_2/10$. In
the direction of increasing area, the lines correpond
to $f = 1, 3, 6$ and 9.}
\protect\label{figcont}
\end{figure}

Figure \ref{figcont} is a contour plot in $\omega_{13}$
and $\chi_{12}$ space of the approximation (\ref{etaBap})
to $\eta_B$. The contours enclose the area when 
$\eta_B > 2 \times 10^{-11}$, for $M_1 =f \times  10^9 $ GeV,
central values of $m_3$ and $m_2$,  $m_1 = m_2/10$,
and are labelled by $f$. Allowing $f > 1$ significantly increases
the available parameter space. This corresponds to
increasing $M_1$ (which should be $\lappeq T_{RH}$),
or increasing $m_3$, which is constrained by
atmospheric neurtino oscillations, or
for  $\chi_{12} \lappeq 0.5$,  to decreasing
$m_2$, which is constrained by solar neutrino experiments
and KamLAND. Perhaps the most palatable way to increase
$f$ is to allow $T_{RH} \sim 10^{10}$ GeV.
The value of $\eta_B$ chosen, $\eta_B = 2 \times 10^{-11}$,
is minimal.  To obtain the CMB favoured  
$\eta_B \simeq 9 \times 10^{-11}$,
would require values of $f$ that were four times larger.

\section{CP violation} 
\label{secCP}

In a previous paper\cite{Davidson:2002em}, we discussed the relation between
the leptonic phases that could be measured at low energy,
and the CP violation required for leptogenesis. 
 We assumed that
$\epsilon$ was large enough, and studied the relative
importance of the neutrino factory
phase $\delta$ and the double beta decay phase
$\phi'$  for leptogenesis. If the 
right-handed neutrinos $\nu_{R1}$ are produced 
{\it non}-thermally, getting $\epsilon$
large enough may not be a significant constraint
(see $e.g.$ \cite{bdps} for a discussion and references).
However, we have seen that it is a challenge when
the $\nu_{R1}$ are produced thermally. So in this section,
we briefly discuss the relative importance of
low-energy phases for thermal leptogenesis---imposing
the constraint that $\epsilon$ is large enough. 

It is well known that there is no linear relation between
the ``leptogenesis phase'' and $\delta$ or $\phi'$
\cite{Branco:2001pq}. That is, the lepton asymmetry
 can be non-zero when $\delta = \phi' = 0$,
and it can be zero when $\delta, \phi' \neq 0$.
To overcome this, we introduced a statistical notion
of ``overlap'' between the leptogenesis phase and
the low energy phases of our parametrisation. The
overlap $O_\delta$ aimed to quantify the relative
importance of the phase $\delta$ for leptogenesis,
assuming that all the low-energy phases were $O(1)$.
In \cite{Davidson:2002em}, we considered the cases
where $W_{13}^2 W_{12}^2 m_3^3 m_2$, or
  $W_{12}^2 W_{11}^2 m_2^3 m_1$, is
the most important term upstairs in $\delta_{HMY}$.
That is, we consider $V_L = 1$, $V_L \sim 1$
and the case of large $V_L$ angles that do not
exactly cancel those in $U$. This occurs
 over most of the parameter space where $\epsilon$ 
could be large enough. However, $\epsilon$ is largest
in the small area of parameter space where $W \sim 1$
and $W_{13}^2 W_{11}^{2*} m_3^3 m_1$
dominates  upstairs in $\delta_{HMY}$. So let
us now consider which low-energy phases are
important for leptogenesis in this case.

Writing the phases explicitely gives
\bea
\epsilon \propto \Im \{ W_{11}^2 W_{13}^{*2} \}
& = & \Im \{ e^{i \phi} [ V_{L11} c_{13}c_{12}
+|V_{L12}| e^{i 2 \varphi_{12}} 
(- c_{23}s_{12}- s_{23}c_{12}s_{13} e^{i \delta} ) \nonumber \\ & &
+|V_{L13}| e^{i 2 \varphi_{13}} 
( s_{23}s_{12}- c_{23}c_{12}s_{13} e^{i \delta} ) ]^2
\times 
[V_{L11} s_{13}e^{-i \delta}
+|V_{L12}| e^{i 2 \varphi_{12}} 
 s_{23}c_{13} \nonumber \\ & &
+|V_{L13}| e^{i 2 \varphi_{13}} 
 c_{23}c_{13}]^2  \}
\eea
where $\varphi_{1j}$ is the phase of $V_{1j}$. 
 $V_{1j} \sim U_{j1}^* e^{i \omega_1}$, because $W \sim 
diag \{ e^{i \omega_1}, e^{i \omega_2},1\}$
\footnote{This constraint on $W$ is what we usually refer
to as $W \sim 1$. Since $V_{11}$ is real, 
 $\omega_1 \simeq - \phi/2$.}.
The ``neutrinoless double beta decay phase'' $\phi'$
is irrelevant for leptogenesis, because it only enters into
$W_{12}$. The phase $\phi$ of $m_1$ will
always be important, because $W_{11} \propto  e^{-i \phi/2}$,
so $\epsilon $ will be  a sum of terms  $\propto
\sin (m \phiŽÂŽÂŽ´ + ...)$. Both the phases
 $\varphi_{12}$ and  $\varphi_{13}$
of $V_{L12}$ and  $V_{L13}$ are likely to have
significant overlap with the leptogenesis phase, because 
$V_L \simeq U^\dagger$ so   the $|V_{L1j}|$
are large.  The ``neutrino factory phase''
$\delta$ always multiplies the CHOOZ angle,
which suppresses its contribution to $\epsilon$.

The three weak-scale phases which 
have significant ``overlap'' with the leptogenesis
phase, in the area of parameter space near $W \sim 1$,
are therefore   $\varphi_{12}$,  $\varphi_{13}$
and $\phi$. This is unfortunate, because although there is some
hope of measuring $\phi'$ and $\delta$, there is 
no foreseeable experiment to determine any of these
three.


\section{Discussion and  summary}
\label{sowhat}

This paper has discussed leptogenesis 
in a minimal model of the SUSY seesaw, with
gravity mediated SUSY breaking and universal soft masses
at  a high scale. It
uses a parametrisation of the model in terms of 
\beq
D_{m_{\nu}},  U,   D_Y,   V_L
\label{list}
\eeq
where $D_{m_{\nu}}$ is the diagonal light majorana
neutrino mass matrix (assumed hierarchical), $U$ is the MNS matrix, 
$ D_Y $ is the diagonal neutrino Yukawa matrix, and
$V_L$ diagonalises $Y_\nu^\dagger Y_\nu$. 
The notation is  briefly defined  in  table,
\ref{tablenotn}.
giving the equation numbers of more detailed definitions.
The angles of
the unitary matrix $V_L$ can be related,
in SUSY models, to the
rates for $\ell_j \rightarrow \ell_i \gamma$
because $Y_\nu^\dagger Y_\nu$ contributes
to the renormalisation group equations for the
slepton masses. This allows the right handed 
neutrino masses and Yukawa couplings
to be expressed as a function
of quantities which could be measured, in principle
or in practise, at the weak scale. This parametrisation
is briefly reviewed in section \ref{revdi1}.

\begin{table}[hbt]
\begin{tabular}{|c|c|c|}
\hline
$ {\bf Y_\nu} = V_R^\dagger D_Y V_L~, y_i $ &
 neutrino Yukawa, eigenvalues & 
\ref{superp}\\
\hline
${\cal M}, M_1 $ &
$\nu_R $ mass matrix, lightest eigenvalue & 
\ref{superp},\ref{M1ap} \\
\hline
$ V_L$ &
$ V_L {\bf Y_\nu}^\dagger {\bf Y_\nu} V_L^\dagger = D_Y^2 $ & 
 \ref{biunitary}\\
\hline
$\varphi_{ij} $ &
 phases of $ V_L $ & 
section \ref{secCP} \\
\hline
$[m_\nu ]~, m_i  $ &
light neutrino mass matrix $ $ & 
 \ref{seesaw} \\
\hline
$U  $ &
MNS matrix $ $ & 
\ref{CKM}  \\
\hline
$\theta_{ij}; \phi,\phi',\delta $ &
 angles; phases of $U $ & 
 \ref{UV}, \ref{Vdef} \\
\hline
$ W $ &
 $V_L U $ & 
\ref{Delta}  \\
\hline
$ \omega_{1j} $ &
  $\simeq $log $[W_{1j}] $ & 
\ref{defom}  \\
\hline
$ \chi_{12} $ &
 $\simeq $log $[U_{12} - W_{12}] $ & 
\ref{defchi}  \\
\hline
$ \eta_{L} $ &
lepton asymmetry $ $ & 
\ref{etaL}  \\
\hline 
$\epsilon  $ &
CP asymmetry $ $ & 
\ref{eps1}, \ref{epsapprox}  \\
\hline
$ \delta_{HMY} $ &
 $\epsilon/ \epsilon_{max}  $ & 
 \ref{bound},\ref{deltaapprox} \\
\hline
$\widetilde{m}_1  $ &
 $\propto \nu_{R1} $ decay rate & 
 \ref{mtilde},\ref{tildekappaap} \\
\hline
$d_1, \tilde{d}_1  $ &
 dilution factor of lepton asymmetry$ $ & 
 \ref{d1},\ref{etaL} \\
\hline
$ \eta_{B} $ &
baryon asymmetry $ $ & 
 \ref{BAU}, \ref{etaBap} \\
\hline
\end{tabular}
\label{tablenotn}
\caption{ Table of notation, with a brief description and
the equation number of a more complete defintion.}
\end{table}

The baryon asymmetry produced in leptogenesis depends on
the number density of right-handed neutrinos
which decay, on $\epsilon \equiv$ the average lepton asymmetry produced
per decay, and on the
survival probability of the asymmetry in the thermal plasma
after it is produced. We consider the ``thermal leptogensis''
scenario, in which the right-handed neutrinos 
are produced by scattering interactions
in the plasma. Non-thermal production mechanisms
are also possible, perhaps even probable, but
depend on additional parameters from the sector
which produces the right-handed neutrinos. Both the
thermally produced $\nu_R$ number density,
and the survival probability of the lepton asymmetry
in the plasma after it is produced,  can be computed
from the reheat temperature of the plasma after
inflation $T_{RH}$, and from the seesaw parameters.
These processes have been carefully studied in
\cite{Plumacher:1997kc,review}.

A convenient
analytic approximation to the numerical
results of \cite{Plumacher:1997kc} is used in
this paper. A single
function $d_1$ (see equation (\ref{d1}))  is defined as the
number density of $\nu_R \times $ the survival
probabilty of the lepton asymmetry once it
is produced. So the baryon to entropy ratio today
is $\eta_B \simeq 8 d_1 \epsilon/23$.  \footnote{where
the 8/23 arises in the transformation of the lepton asymmetry
into a baryon asymmetry by the electroweak $B+L$ violating
processes.}

The $CP$ asymmetry produced in
the decay of the lightest $\nu_R$ is
bounded above (for hierarchical $M_i$ and $m_j$):
\beq
\epsilon < \frac{3 M_1 m_3}{8 \pi v_u^2}
\label{bdinsum}
\eeq
where $M_1$ is the mass of the $\nu_{R1}$ and 
$m_3 = \sqrt{\Delta m^2_{atm}}$. Since $d_1$ cannot
exceed $45/(2 \pi^4 g_*)$ for thermally
produced $\nu_{R1}$, obtaining $\eta_B > 3 \times 10^{-11}$
requires $M_1 > 3 \times 10^8$ GeV. 

In section \ref{secapprox}, approximate
analytic  formulae
for the lightest $\nu_R$ mass $M_1$, for the CP
asymmetry $\epsilon$ and for the baryon asymmetry $\eta_B$,
are given in terms of our weak-scale parameters.
These approximations are valid for  hierarchical $M_i$ and
neutrino Yukawa eigenvalues $y_j$. 
The baryon asymmetry can be written as a function
\beq
\eta_B (m_2,m_3,\theta_{23},\theta_{12};\theta_{13},
V_{L12},V_{L13}, y_1,m_1, phases)
\eeq
where ``known'' low-energy parameters precede
the semi-colon.  We concentrate on the dependence
of $\eta_B$ on real parameters, assuming that
the phases can be chosen to maximise the asymmetry. 

It is interesting that the baryon asymmetry only depends on
5 of the 8 unknown real parameters in eqn (\ref{list}). 
Two of these, 
$\theta_{13}$ and $V_{L13}$, are possibly measurable;
the constraints that thermal leptogenesis
imposes on them  will be discussed later for
different areas of parameter space.
On the other hand, there are
no foreseen experiments that could  determine
$y_1$, $m_1$, and $V_{L12}$.
$V_{L12}$ is included in
the discussion with $\theta_{13}$ and $V_{L13}$, because
 these three unknowns can be exchanged for the 
12 and 23 elements of $W = V_L U$. This
simplifies expressions and is
convenient for plotting.
The dependence of $\eta_B$ on
$m_1$ is  subtle, comparatively unimportant, and
discussed in an Appendix. 
The lightest right-handed
neutrino mass $M_1$, and therefore
the baryon asymmetry, is proportional to  $y_1^2$.
So   $y_1^2$ is exchanged for $M_1$. This is a peculiar exchange---
why 
do we want to use a GUT-scale
mass as input in our weak-scale parametrisation?
The off-diagonal elements of $V_L$, and the $y_i$, are
related to
lepton flavour violating off-diagonal slepton mass
matrix entries (to which processes like $\ell_j
\rightarrow \ell_i \gamma$ are sensitive), and to
slepton mass differences.  The smallest neutrino Yukawa
$y_1$ makes negligeable contributions to both
these effects. However,
$M_1 > 3 \times 10^8$ GeV  is required for
thermal leptogenesis to have any hope of working, 
and if SUSY is discovered,
the sparticle spectrum could give some indication of
the gravitino mass, and therefore the allowed reheat temperature
$T_{RH}> M_1$. So we ``determine''
$y_1$ by requiring that thermal leptogenesis
$could$ produce a large enough asymmetry: 
$ 3 \times 10^8$ GeV $ < M_1  < T_{RH}$. 
Then we study
the  requirements on the remaining parameters such that 
the asymmetry  $is$ large enough.
These additional requirements may have observable consequences.

The analytic 
formulae of section \ref{secapprox}
are simple and compact, but 
nonetheless difficult to visualise. 
The asymmetry depends on the first row of the matrix
$W$, so for
qualitative understanding,
we show 3-dimensional figures of leptogenesis parameters as a function of
$\omega_{13} \simeq $ log$W_{13}$, and 
$\omega_{12} \simeq $ log$W_{12}$. 
Logarithmic measure is reasonable for
unmeasured but observable matrix
elements---which the $W_{1j}$ are {\it not}.
For small angles in $V_L$ (see section \ref{when}
for a general discussion), the $W_{1j}$ can be
related to the more physical matrix elements 
$V_{L1k}$ and $U_{ij}$:
$W_{13} \sim \theta_{13} + V_{L12} + V_{L13}$,
$W_{12} \sim \sin \theta_{sol} + V_{L12} + V_{L13}$.
To present the area of parameter space
where leptogenesis works with a more
physical measure, we therefore  plot the
baryon asymmetry as a function of 
$\omega_{13}$ and $\chi_{12}\sim $ log$ [  V_{L12} + V_{L13}]$
in figures \ref{figetaB} and \ref{figcont}.

 We define thermal leptogenesis
to ``work'' if it can produce  $\eta_B \gappeq
3 \times 10^{-11}$, as required by Big Bang Nucleosynthesis.
For $M_1 \simeq 10^9$ GeV
(consistent with the canonical gravitino bound $T_{RH} \sim 10^9$ GeV), 
leptogenesis {\it can} work:
there is a limited parameter
space where the upper bound on $\epsilon$
is almost saturated, $and$ $d_1$ is close to maximal.
This can be seen in figure \ref{figcont},
where the baryon asymmetry is large
enough inside the contours, which are 
 labelled by $f$, where $M_1 = f \times 10^{9}$ GeV.
Increasing
$T_{RH}$ (and thereby the allowed $M_1$)  enlarges
the parameter space where thermal leptogenesis
works.

We now come to the aim of the paper---
what are the weak scale foot prints of thermal leptogenesis?
What parameter values  must be observed,
if thermal leptogenesis works in an MSUGRA model?

Suppose first that $f \simeq 1$, which corresponds
to   $M_1 \sim 10^9$ GeV for central values of the light
neutrino masses. Thermal leptogenesis works in the
 area of figure \ref{figcont} at $\omega_{13} \sim  -2$
 and  $\chi_{12} \sim -0.5$. This small area of parameter
space is discussed in section \ref{Wsim1}, and occurs
if  $W = V_L U \sim 1$. The phenomenological consequences
of this area of parameter space are unambiguous: the branching ratio
of $\tau \rightarrow \mu \gamma$, or $\tau \rightarrow e \gamma$,
should be large. More concretely,
at least one of  $V_{L32}$ or $V_{L31}$ is $O(1/\sqrt{2})$,
so according to the estimates of table \ref{tabmueg},
$BR(\tau \rightarrow \ell \gamma) \gappeq 10^{-8}$.

From a theoretical model building perspective, this
area of parameter space corresponds to
the neutrino Yukawa and light  mass matrices
$Y_\nu$ and $[m_\nu]$ being almost simultaneously
diagonalisable. The large leptonic
mixing angles arise in the rotation from this
basis to the one where the charged lepton Yukawa
matrix $Y_e$ is diagonal.

The baryon asymmetry is largest at this point for two reasons.
The $\nu_R$ decay rate (eqn (\ref{mtilde})) is small,
so lepton number violation is slow after the asymmetry is produced,
and more of the asymmetry survives. Secondly, the asymmetry
produced is almost maximal; it comes within a factor
of $O(1)$ of the upper bound eqn (\ref{bdinsum}). This is discussed
after eqn (\ref{deltaapprox}). 

The {\it area} of this parameter space, where $\eta_B$ is
largest, depends on
the smallest neutrino mass $m_1$. The plots are made
with $m_1 = m_2/10$; the area shrinks as $m_1$ decreases.
This peak in $\eta_B$ only exists for $m_1 \neq 0$. It is
interesting that the CP asymmetry and the low-energy
footprints of this area of parameter space are {\it independent}
of $m_1$. However,
the number density of $\nu_R$ (and therefore the baryon asymmetry)
decreases for $m_1 \lappeq 10^{-5}$ eV, and our approximation
fails.  See the Appendix  for a discussion.

In brief, if  a sparticle spectrum 
consistent with gravity mediated SUSY breaking is measured, with
a gravitino mass of $m_{3/2} \sim 100$ GeV ($T_{RH} \sim 10^9$
GeV), then  $\tau \rightarrow \mu \gamma$  or  $\tau \rightarrow e \gamma$
must be observable for thermal leptogenesis
to work.

Now consider the enlarged parameter space allowed
by $ M_1/(10^9$ GeV) $\equiv f> 1$
in figure \ref{figcont}:
thermal leptogenesis works for $W_{13}
\sim m_2/m_3$ and pretty much all values of
$W_{12}$. 
This sets one constraint on the three
''physical'' matrix elements  $\sin \theta_{13}$,
 $V_{L13}$ and $V_{L12}$. As discussed in
sections \ref{VL=1} and \ref{VLneq1}, it can be
satisfied if any one of the angles is $O(m_2/m_3)$.
These possibilities have different weak-scale implications.

If $V_L$ has small angles like the CKM matrix, 
 $V_{L13}, V_{L12} \ll 0.1$, then leptogenesis
requires that the CHOOZ angle  $\theta_{13} \simeq 0.04$,
which is close to its current experimental bound. 
This  implies that
the baryon asymmetry is determined by parameters
which can be measured in the neutrino
sector
\footnote{Caveat: $\eta_B \propto  M_1 m_3/m_2$
in this case, and we ``determine''
 $M_1 \simeq 6 \times 10^{9}$ GeV by requiring $\eta_B$ large
enough. If $m_3$ ($m_2$) is larger (smaller)
than the current best-fit values, $\eta_B$ increases.}: 
$m_3$, $m_2$,  $\theta_{13}$ and the phases
$\delta$ and $\phi'$.

If the CHOOZ angle $\theta_{13} \ll 0.1$, then it is
still possible to sit on the $W_{13} \sim 0.04$ ridge,
by having $V_{L13}$, or $V_{L12} \sim 0.04$.
The former angle is related to  $\tau \rightarrow e \gamma$,
and could perhaps be measured in this process.
 Unfortunately, $V_{L12}$ appears
in the RGEs multiplying $y_2^2$, the middle Yukawa eigenvalue,
so has no observable consequences in the slepton
mass matrix. So thermal leptogenesis can ``work''
when
$\theta_{13}$ and $BR(\ell_j \rightarrow \ell_i \gamma)$
are  unobservably small. 

The matrix elements  $V_{L13}$, and $V_{L12}$   are
small along most of the $W_{13}$ ridge
currently under  discussion. Many texture models
occupy this area of parameters space, where
the CKM-like matrix $V_L$ (between the bases where
$Y_\nu$ and $Y_e$ are diagonal) has small angles. The
large angles of the MNS matrix then
arise from the majorana structure of $Y^T {\cal M}^{-1} Y$.

The baryon asymmetry is larger along the ridge
than in the rest of parameter space, because the washout
is moderate---$\tilde{m} \sim m_2$---and because
the CP asymmetry $\epsilon$ approaches its upper
bound (\ref{bdinsum}).
Notice that the baryon asymmetry on this ridge is
independent of $m_1$---the solution remains as $m_1 \rightarrow 0$.
 As discussed
after eqn (\ref{deltaapprox}), there are two limits
in which
$\epsilon$ is maximal: the ridge
 where $W_{13} \sim m_2/m_3$,
and the previously discussed peak.
There is an orthogonal ridge in figure \ref{figcont}, at $\chi_{12} \sim -0.5$
($W_{12} \sim 0.1$), where $\delta_{HMY} \lappeq m_2/m_3$,
but washout is minimised. It has the same observable
footprints as the peak.

Until now we have considered {\it if} thermal leptogenesis
works, then {\it what should we see at low energy?} 
Allowing $T_{RH } \sim 10^{10}$ GeV, it seems just about
all phenomenology is consistent with thermal leptogenesis:
large $\tau \rightarrow \mu \gamma$, $\theta_{13} \sim 0.04$,
observable  $\tau \rightarrow e \gamma$, nothing observable
at all...So now consider the inverse question:
are there weak-scale observations that can 
rule out thermal leptogenesis?
The previous discussion
is vague because SUSY has not been discovered.
Clearly thermal leptogenesis does not work if $W_{13}$ is too big
or too small. Since  all the terms
which contribute to $W_{13}$ cannot be
measured, no experimental lower bound can be set.
 However, one could tell that  $W_{13}$
is too $large$, for instance if 
large $V_{L13}$ 
($\tau \rightarrow e \gamma$)
is measured\footnote{ If there are additional sources
of flavour violation in the slepton masses, ({\it e.g}
non-universal soft masses) this does not work.}.

The analysis of this paper relies crucially on the assumption that the
$\nu_R$ are produced thermally. A larger
number density of the lightest $\nu_R$, $n_{\nu_R}/s$,
could be produced non-thermally, so a 
large enough baryon asymmetry could be produced with
a smaller $\epsilon$. This  would enlarge the available parameter
space. Furthermore, if the $\nu_R$ are produced
non-thermally, they could be $\nu_{R2}$ or  $\nu_{R3}$,
making the formulae for  $\epsilon$ inapplicable.

In summary,  we study the baryon asymmetry resulting
from the decay of the lightest right-handed neutrino
$\nu_{R_1}$, assuming the $\nu_{R_1}$s are produced
thermally. We present compact analytic approximations
for the quantities relevant to thermal leptogenesis,
in terms of the light neutrino masses, the MNS matrix,
the smallest eigenvalue of the neutrino Yukawa matrix
$Y_\nu$, and the matrix $V_L$ which diagonalises
$Y_\nu$ on its SU(2) doublet indices. In the MSUGRA scenario,
we can trade these parameters for the neutrino and
sneutrino mass matrices ($m_\nu$ and $m_{\snu}^2$), or more usefully, 
for  $m_\nu$, for the branching ratios of lepton
flavour violating decays $\ell_j \rightarrow \ell_i \gamma$,
and for the lightest right-handed neutrino mass $M_1 \lappeq T_{RH}$.
We find a small area of parameter space where a large enough
baryon asymmetry is generated for $T_{RH} \sim 10^9$ GeV.
It corresponds to large off-diagonal elements in  $m_{\snu}^2$,
and therefore observable $\tau \rightarrow e \gamma$. 
For $T_{RH} \sim 10^{10}$ GeV, leptogenesis can also work
for smaller off-diagonal elements in
$m_{\snu}^2$.

\subsection*{Acknowledgements}
Thanks to Oxford where we started this, and 
to Valencia, for a warm and sunny 
welcome when it was being completed.
I am grateful to  Marco Peloso
for encouragement and asking the right questions, and  
to Michael Pl\"umacher for many discussions, comments and
for careful reading of the manuscript. I 
particularily thank   Alejandro
Ibarra for 
innumerable  productive discussions and important
contributions.

\subsection*{Note added}
After this work was completed, related analyses
\cite{new} appeared. 

\section{Appendix:the approximation and plots}

In this Appendix, 
 the lightest eigenvalue and 
corresponding eigenvector of
$\cal M$ are estimated, using  an approximation
borrowed from diagonalising  neutrino mass matrices
in R-parity violating theories.

 It is first convenient to
  scale  some powers of the smallest
Yukawa out of 
the hermitian matrix $ {\cal M}^{-1 \dagger} {\cal M}^{-1}$:
\beq
v_u^4{\cal M}^{-1 \dagger}{\cal M}^{-1} 
=  D_Y^{-1} V_L [m_\nu]^{\dagger} V_L^TD_Y^{-2} V_L^* 
[m_\nu] V_L^{\dagger} D_Y^{-1}
\equiv \frac{ \Lambda}{y_1^4} ~~.
\eeq
This can be written more compactly as
\bea
 \frac{\Lambda}{y_1^4} & =& 
 {\bf  D_Y^{-1}} \Delta^{\dagger} {\bf  D_Y}^{-2}  \Delta {\bf  D_Y^{-1}} ~~,
\label{omega}
\eea
by defining 
\beq
\Delta = V_L^* [m_\nu] V^{\dagger}_L 
= V_L^* U^* D_{m_\nu} U^{\dagger} V_L^{\dagger}
\equiv W^* D_{m_\nu} W^{\dagger} ~~,
\label{Delta}
\eeq
where the matrix $W=V_L U$ is 
the rotation from the basis where the $\nu_L$ masses
are diagonal to the basis where the neutrino Yukawa  matrix
${\bf Y^{\dagger}_{\nu}} {\bf Y_\nu}$ is diagonal.

\begin{figure}[ht]
\includegraphics[width=7cm]{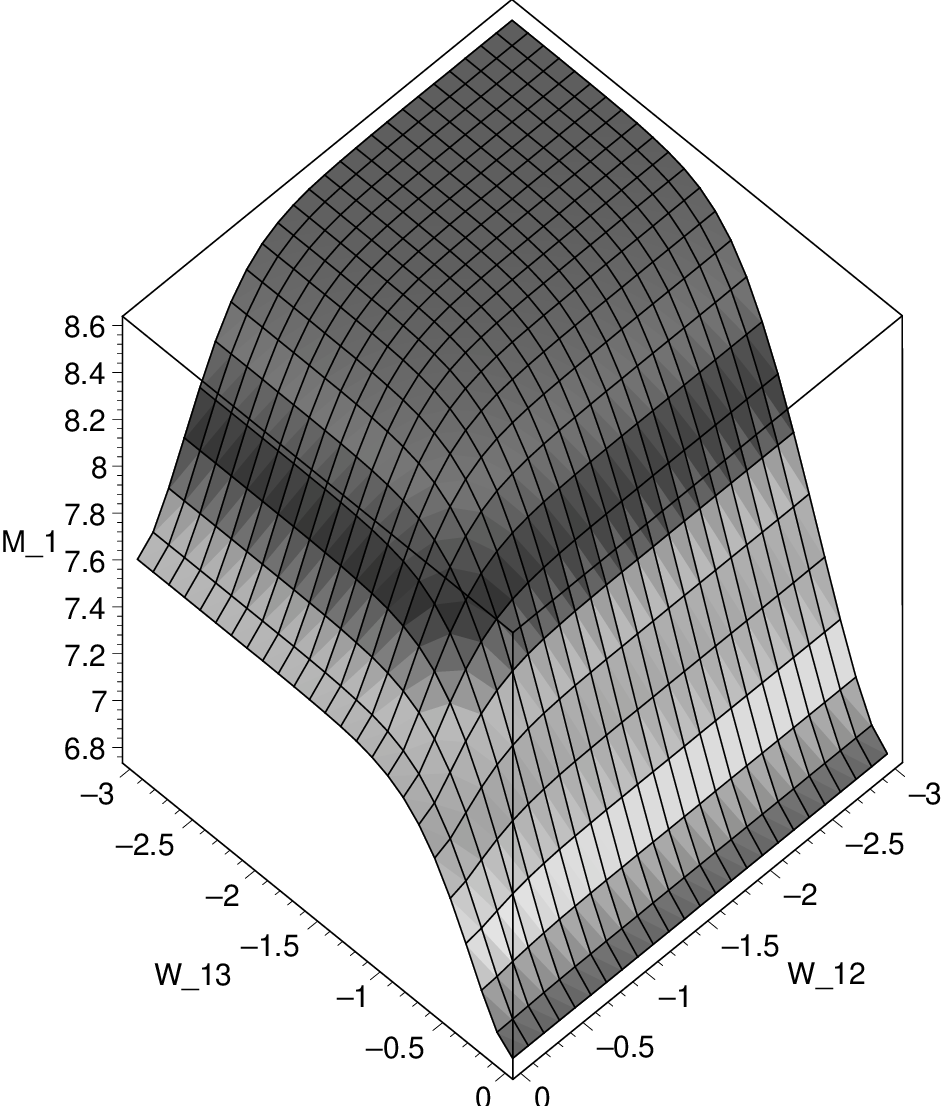}
\includegraphics[width=7cm]{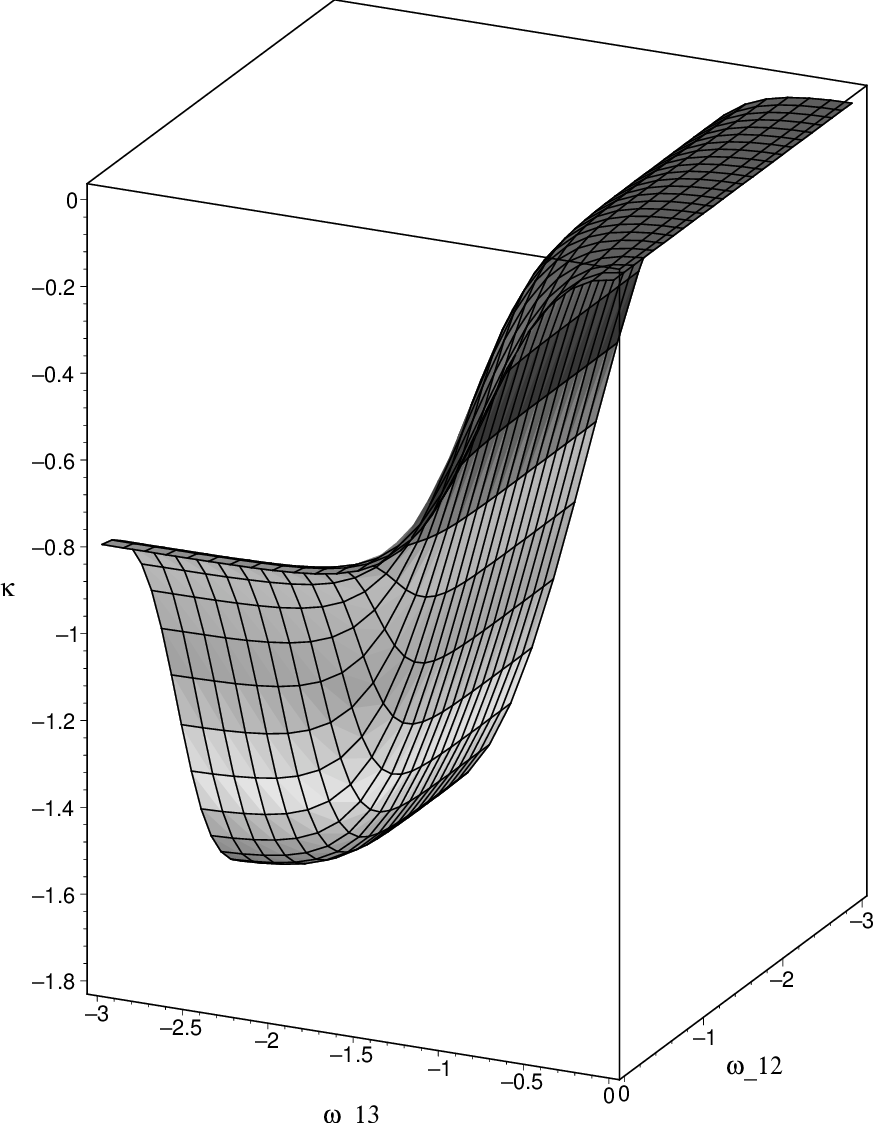}
\caption{On the LHS, $\log  [ M_1/{\rm GeV}]$ as a function of 
$\omega_{12} \simeq log [ W_{12} ]$ and $\omega_{13} \simeq log [ W_{13} ]$.
On the RHS,   $\kappa = log_{10}(\tilde{m}_1/m_3)$. 
Recall we need $\kappa \lappeq -1.4$ to maximise the asymmetry.
These plots are for central
values of the neutrino masses:
$y_1 = 10^{-4}, m_2 = 8.2 \times 10^{-3}$ eV, and $ 
m_3 = 5.2 \times 10^{-2}$ eV.}
\protect\label{figM1}
\end{figure}

The matrix $\Lambda$ can be written 
\bea
[\Lambda]_{ij} & =& 
 ( \vec{\lambda}_i) \cdot  (  \vec{\lambda}_j^\dagger) = 
 \sum_k ( {\lambda}_i)_k  ( {\lambda}_j^*)_k ~~,
\label{omega2}
\eea
where 
\beq
\label{lamvec}
\vec{\lambda}_i \equiv  \frac{y_1}{y_i} \left(\begin{array}{c}
\Delta_{1i}^* \\
y_1\Delta_{2i}^* /{y_2} \\
y_1\Delta_{3i}^* /{y_3}
\end{array} \right)  ~~.
\eeq
If the  hierarchy in the $y_i$  is steeper than in
the $m_j$, and/or that the angles in $W$ are large,
then
\beq
|\vec{\lambda}_1|^2 \gg |\vec{\lambda}_2|^2,|\vec{\lambda}_3|^2
\eeq
so  the largest eigenvalue of $\Lambda$ 
($ = v_u^4 y_1^4/|M_1|^2$) 
is
\beq
|M_1| \simeq \frac{y_1^2v_u^2 }{\sqrt{|\lambda_1|^2}}
\simeq  \frac{y_1^2v_u^2 }{|\Delta_{11}| }
= \frac{y_1^2v_u^2 }{|W^2_{1j} m_j| }
\label{M1apap}
\eeq
with associated eingevector (normalised $\vec{\lambda}_1$):
\beq
\label{eigenvec1}
\hat{\lambda}_1 \simeq   \left(\begin{array}{c}
\Delta_{11}^* \\
y_1\Delta_{21}^* /{y_2} \\
y_1\Delta_{31}^* /{y_3}
\end{array} \right) \times\frac{1}{ \Delta_{11}^*}
\eeq
In figure \ref{figM1}, $M_1$ is plotted as a function of
$\omega_{12} \simeq \log W_{12} $ and 
$\omega_{13} \simeq \log W_{13}$, for  $y_1 = 10^{-4}$, and
using  the central values of $m_i$
listed after eqn (\ref{CKM}).
The precise definition is 
\bea
  W_{12}  = 
\cos \theta_{W13} \sin \theta_{W12},  &&
W_{13} =  \sin \theta_{W13} \nonumber \\
 with ~~
\theta_{W1j}& =& 10^{ \omega_{1j}} \pi/2 ~~.
\label{defom}
\eea
 This Appendix contains  many three
dimensional plots of functions
that will enter into the equation for
the baryon asymmetry. The aim of these
figures is to give a qualitative
impression;  quantitatively clearer contour
plots of the baryon asymmetry are in the body of the
paper. In figure \ref{figM1},
there are three limiting values for $M_1$, 
corresponding to $M_1 \simeq y_1^2v_u^2/m_i$:
$M_1 \rightarrow  y_1^2v_u^2/m_3$ when $W_{13} \rightarrow 1$,
$M_1 \rightarrow  y_1^2v_u^2/m_1$ when $W_{13}, W_{12} \rightarrow 0$,
and   
$M_1 \rightarrow  y_1^2v_u^2/m_2$ when $W_{13}<m_2/m_3 $, 
$W_{12} \rightarrow 1$.

The $\nu_R$ decay rate can be evaluated with  the eigenvector
(\ref{eigenvec1}), which gives eqn 
(\ref{tildekappaap}). 
$\tilde{m}_1$ has three limits---$m_1, m_2, m_3$---depending ono
the values of $W_{1j}$. 
The logarithm of $\tilde{m}_1/m_3$ is plotted on the RHS
of figure \ref{figM1}. $\tilde{m}_1$ must be in the range given
after eq. (\ref{mtilde}), 
which implies log $(\tilde{m}_1/m_3) \lappeq -1.4$.

\begin{figure}[ht]
\includegraphics[width=7cm]{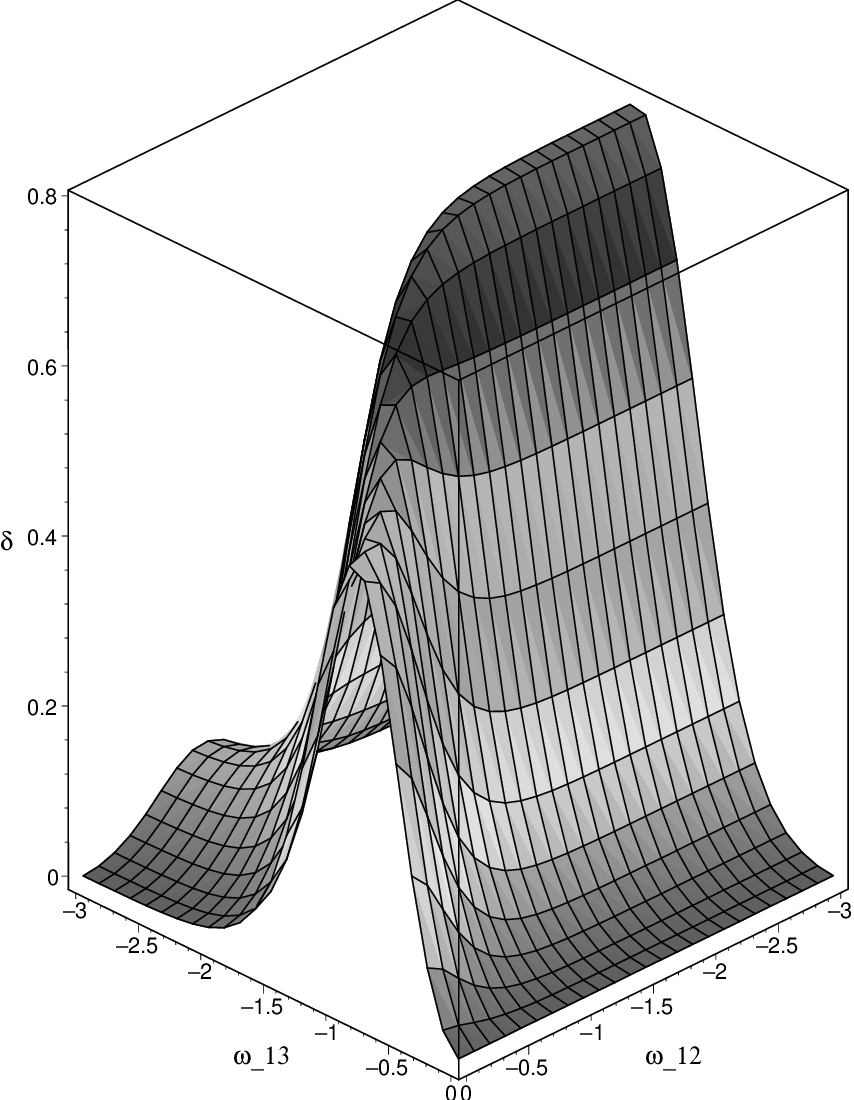}
\includegraphics[width=7cm]{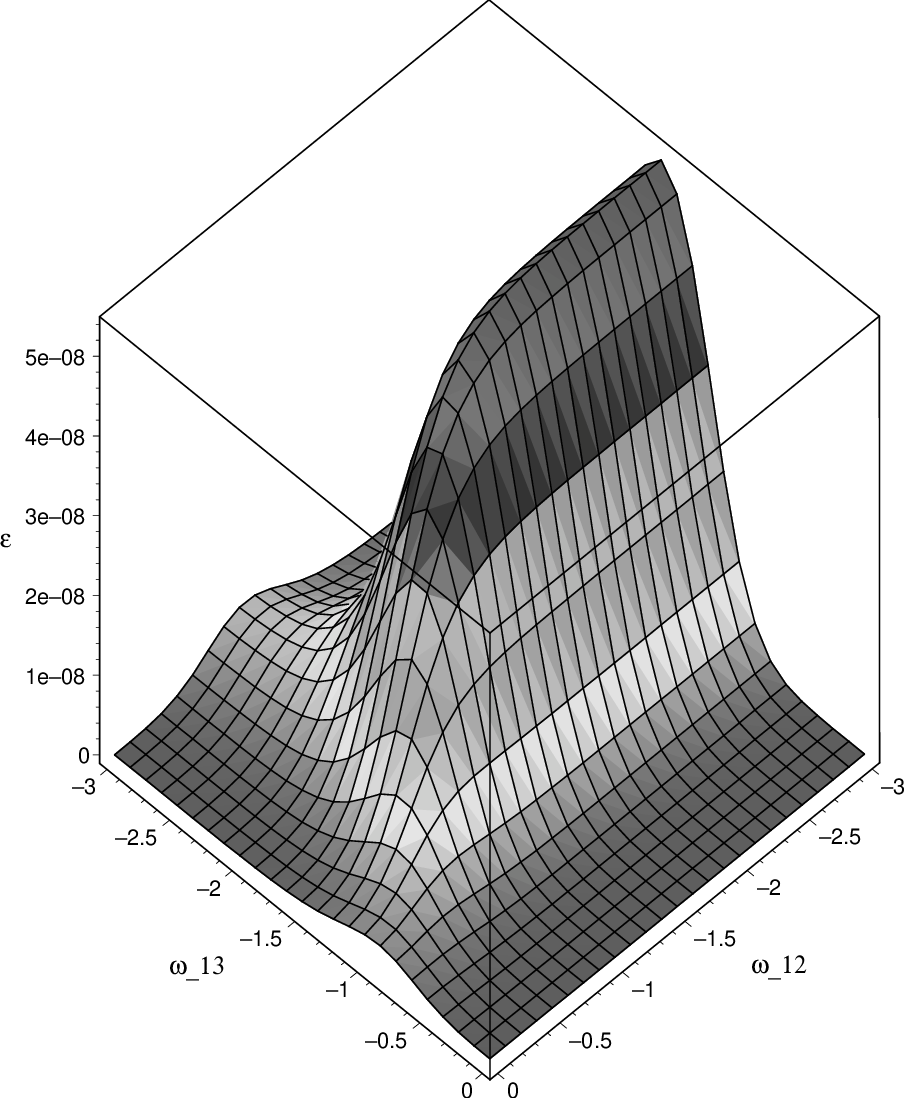}
\caption{
On the LHS (RHS), $\delta_{HMY}$
($\epsilon$) as a function of 
$\omega_{12} \simeq log [ W_{12}]$ 
and $\omega_{13} \equiv log [ W_{13} ]$. 
On the RHS, $M_1$ is taken as a function
of $y_1 = 10^{-4}$ and other inputs.
Both plots are for central values of the neutrino
masses. }
\protect\label{figeps3d}
\end{figure}

Finally, the CP asymmetry $\epsilon$,  eqn. (\ref{eps1}),
can be evaluated  with  the eigenvector (\ref{eigenvec1}) to obtain
\beq
\epsilon \simeq 
- \frac{3  \Lambda_{11}^2}{8 \pi [\Lambda D_Y^2 \Lambda]_{11} }
{\rm Im} \left\{ 
\frac{[\Lambda D_Y \Delta^\dagger D_Y \Lambda^T]_{11}}
{[\Lambda  D_Y^{-1}  \Delta^\dagger  D_Y^{-1}  \Lambda^T]_{11}}\right\}
=
\frac{3 y_1^2}{8 \pi \sum_j |W_{1j}|^2  m_{\nu_j}^2} 
 {\rm Im} \left\{ \frac{ \sum_k W_{1k}^{2} m_{\nu_k}^3 }
{ \sum_n W_{1n}^{2} m_{\nu_n} } \right\} ~~,
\label{epsapprox}
\eeq
where terms of order  $y_1/y_2$ and $y_1/y_3$ have been dropped.
$ \delta_{HMY} \propto \epsilon_1/ M_1$ is given in eqn
(\ref{deltaapprox}), and plotted on the LHS of figure
\ref{figeps3d}. $\epsilon_1$ 
is plotted on the RHS; it peaks  on the ridge
of eqn (\ref{aaa})  because this is where the larger
values of $M_1$ and $\delta_{HMY}$ overlap.

The results in the remainder of the paper are based
on the analytic approximations of this section. How
reliable are these equations?
The eqn (\ref{omega}) for $ \Lambda = y_1^4 [{\cal M}{\cal M}^\dagger]^{-1}$
 is exact, but
the formula we use for $\epsilon_1$ assumes hierarchical
$M_i$, so it is consistent to assume this in solving
for the eigenvalues and eigenvectors of $\Lambda$.
The approximation is that the first column (or row)
of $y_1^2  D_Y^{-1} \cdot \Delta \cdot D_Y^{-1}$ 
is the lightest  eigenvalue, multiplying
its eigenvector. 
It breaks down if the elements of the second or third row/column
become of order $\Delta_{11}$, as one can see
by writing the eigenvector in a basis rotated by a small
angle from the eigenbasis.   $y_1 \Delta_{12}/y_2,  
y_1\Delta_{13}/y_3 \simeq \Delta_{11}$ 
could occur if
\begin{enumerate}
\item  the $M_i$ were of similar magnitude, rather than
hierarchical.  This is  ``unlikely'',  because the hierarchy in
$D_Y$ is much steeper than in $[m_\nu]$.
\item $m_1$ too small---if $m_1/m_2,m_1/m_3 < y_2^2$, then
the terms being kept are smaller than the  neglected ones.
This is discussed in an appendix. 
\end{enumerate}

\section{Appendix: $m_1 \ll m_2/10$ }
\label{appB}

 In this paper,
we assumed a hierarchical spectrum for the light neutrino masses:
$\Delta m^2_{atm} = m_3^2$, $\Delta m^2_{sol} = m_2^2$,
so the smallest neutrino mass $m_1$ is unlikely to
be measured with anticipated data.
However, it enters our formulae for the baryon asymmetry,
as does the smallest Yukawa $y_1$.
In the body of the paper, we fixed 
$m_1 = m_2/10$, and determined $y_1$ as a
function of $M_1$ and our weak scale parameters,
by requiring $M_1$ to be in the range allowed by
leptogenesis.  In this Appendix, 
we discuss the dependence of our results on
$m_1$. For most values of
$W_{12}$ and $W_{13}$, $m_1$ is irrelevant
because $|W_{11}|^2 m_1 \ll |W_{12}|^2 m_2,|W_{13}|^2 m_3$.
However, $m_1$ cannot be dropped from our analytic expressions,
 for $W$ close to the identity.
This is the area of parameter space where $\eta_B$ is maximal;
the remainder of the Appendix is restricted to
this area of parameter space.
 We are interested in how $\eta_B$ scales with
$m_1$, and  in whether
 our analytic approximation is  still valid.

The maximum value  of  $\epsilon$,
eqn (\ref{bdinsum}), 
is independent of $m_1$, if $M_1$ is independent
of $m_1$. We have fixed $M_1 \simeq T_{RH}$,
which determines $y_1^2$ as a function of
$W_{1n}^2 m_n \sim m_1$. So varying
$m_1$ allows $y_1$ to vary:
$m_1 \sim m_2/100$ would allow
leptogenesis to work for $y_1 \sim h_u$,
which could be theoretically attractive.

As discussed after eqn (\ref{deltaapprox}),
$\epsilon$ approaches its upper bound 
(equivalently $\delta_{HMY} \sim 1$)
on the peak
in figure \ref{figeps3d},   where 
$m_1^2/m_3^2  \sim W_{13}^2 $ and  
$ W_{12}^2 <m_1^2/m_2^2$.  As $m_1$ decreases,
the {\it area} in $W_{12},  W_{13}$ space
where  $\delta_{HMY} \sim 1$ decreases, but the
maximum value is unchanged. The numerical values of 
$W_{12}$ and $  W_{13}$ where the maximum is reached
will also decrease, making this parameter space
increasingly ``fine-tuned'' ($W$ very close to
the identity is unlikely to be stable under renormalisation
group running).

The $\nu_R$ decay rate must have values in the range given after eqn
(\ref{mtilde}), to ensure $\eta_B$ as 
large as possible. We first concern ourselves
with the upper bound: $\tilde{m}_1 < 3 \times 10^{-3}$ eV. 
 If $ W_{1n}^2 m_n^2  \sim  W_{13}^2  m_3^2$
 and $ W_{1j}^2 m_j  \sim  W_{11}^2  m_1$, as required
to maximise  $\delta_{HMY}$,
 then from eqn
(\ref{tildekappaap}),  $\tilde{m}_1 \sim  W_{13}^2  m_3^2/ m_1$.
 To maximise $\eta_B \sim \delta_{HMY}/\tilde{m}_1 $,
requires  $W_{13}^2  \simeq m_1^2/m_3^2$, so that the
decay rate is slow enough, but $ \delta_{HMY}$ is still
$O(1)$. 
So as $m_1$ decreases from $m_2/10$ to $10^{-3}m_2$, the
area of  the peak on the RHS of figure \ref{figetaB} will
shrink, but the height is unchanged.

For smaller values of $m_1$, the asymmetry
will decrease. This is because 
$\tilde{m}_1$ is small, so $\nu_R$ production in the plasma is
inefficient (see \cite{review}).

It is straightforward to check that 
 our analytic approximation holds, in
the shrinking area of parameter space where
$W_{13}  \simeq m_1/m_3$ and $W_{12}  < m_1/m_2$,
provided that $y_1^2/m_1 \ll y_2^2/m_2$.
This is the condition that  $y_1^2v_u^2/m_1$ is the lightest
 $\nu_R$ mass. So the analytic approximation fails
as $m_1$ approaches $ \frac{y_1^2}{y_2^2} m_2$.

Finally, the low energy prediction of the peak are
independent of $m_1$, because they follow from requiring
that $W \sim 1$. As $m_1$ decreases, $W$ must approach
the identity more and more closely, so $V_L$ becomes more precisely
$U^\dagger$. But whether $V_L \sim U^\dagger$, or 
$V_L = U^\dagger$, the expectation  remains that $\tau
\rightarrow \mu \gamma$ or $\tau
\rightarrow e \gamma$ should be observable.

So in summary, the magnitude of the baryon
asymmetry on the peak of figure \ref{figetaB}
is independent of $m_1$ for 
$10^{-3}m_2 < m_1 <m_2/10$.
As $m_1$ decreases, the location of the peak
shifts to smaller  $W_{13}$,
and its area will shrink.
We cannot say anything for  $m_1 <10^{-3}m_2$: our analytic formulae
indicate that $\eta_B$ will decrease, but the approximation
they are based on is unreliable.

\end{document}